# Memory-Generated Transport Geometry: Curvature, Holonomy, and Irreversibility

Mounir Kassmi
University of Tunis El Manar: Campus Universitaire Farhat Hached B.P. n 94-Rommana, 1068 Tunis, Tunisia, Faculty of Sciences of Tunis

Email: mounirkassmi60@gmail.com

## Abstract

Memory is traditionally incorporated into transport theory as a constitutive correction acting on an already prescribed kinematic structure. Here we develop a different framework in which finite memory itself generates the geometry of transport. By reconstructing deformation from causal transport histories, the instantaneous velocity gradient is replaced by a memory-dependent transport connection whose ordered evolution gives rise to noncommutativity, curvature, and holonomy in transport-history space.

We show that finite memory generates a nonvanishing geometric contribution to transport even in time-periodic, irrotational flows, providing a purely kinematic mechanism for irreversible Lagrangian transport without invoking vorticity, constitutive nonlinearities, stochastic forcing, or explicit symmetry breaking. We introduce an intrinsic curvature invariant, independent of the transport representation that measures the accumulated geometric structure generated by transport history.

The framework predicts universal scaling governed by the dimensionless parameter ($\omega\tau_m$), identifies a characteristic memory scale ($\tau_c$) separating rapid geometric accumulation from asymptotic saturation, and reveals a monotonically decreasing memory susceptibility with globally concave accumulation dynamics. Numerical simulations confirm these predictions and show that geometric irreversibility emerges through progressive curvature accumulation rather than resonance-driven amplification. These results establish finite memory as a generator of an intrinsic geometric structure rather than merely a modifier of dynamical evolution, revealing causal history as the microscopic origin of curvature, holonomy, and irreversible transport across a broad class of non-Markovian systems.

## 1 Introduction

Transport underlies the evolution of virtually all nonequilibrium physical systems, from classical fluids and soft condensed matter to active media, plasmas, biological systems, and quantum materials. Since the development of classical continuum mechanics, transport has been described in terms of local kinematic quantities evaluated at the present time, with the instantaneous velocity gradient serving as the fundamental descriptor of deformation. This local viewpoint, established through the continuum formulation of fluid mechanics by Landau and Lifshitz, Batchelor, and the general framework of nonlinear continuum mechanics developed by Truesdell and Noll, has provided an extraordinarily successful description of transport over a vast range of physical systems [1–3].

The success of this local description, however, relies on the assumption that the present state contains all the information required to determine subsequent evolution. This assumption, however, is no longer adequate for a broad class of physical systems. Polymeric fluids, viscoelastic materials, biological tissues, active matter, glasses, and quantum environments exhibit finite memory, whereby the present dynamics depend on a finite interval of previous evolution. The recognition of such hereditary effects can be traced back to Maxwell's pioneering relaxation model [4, 5], which was subsequently generalized through the constitutive theories of Oldroyd [6] and later incorporated into the modern theory of complex fluids by Bird, Armstrong, and Hassager [7]. Since then, finite memory has become a central ingredient in the theoretical description of transport across a broad spectrum of nonequilibrium systems. A second major development emerged from nonequilibrium statistical mechanics. The projection-operator formalism introduced by Mori [8] and the generalized Langevin approach developed by Zwanzig [9, 10] established memory as a fundamental mechanism governing irreversible dynamics through temporal correlations extending beyond the instantaneous state of the system. Over the past decades, this perspective has expanded considerably through generalized master equations [20], open quantum systems [11, 15], anomalous diffusion [16, 17, 21], collision and transport models [18, 19, 24], coarse-grained molecular dynamics [25], nonequilibrium statistical mechanics [26, 27], emergent memory in active matter [28], and generalized hydrodynamics [36].

Taken as a whole, these studies position finite memory as one of the defining features of modern nonequilibrium transport. Beneath their striking diversity, they converge on a single conceptual bedrock.

Memory enters the description through constitutive equations, relaxation kernels, generalized friction operators, or temporal correlation functions, thereby governing the dynamical evolution of the system. While the geometric framework remains prescribed, however, it is assumed from the outset. Geometry provides the stage on which transport evolves, whereas memory modifies only the dynamics occurring upon that stage. Consequently, finite memory has traditionally been regarded as a dynamical ingredient rather than as a possible origin of geometric structure.

Meanwhile, an independent line of research has progressively revealed the central role of geometry throughout modern physics. In parallel with the development of memory theories, geometry has progressively emerged as a unifying language across many areas of modern physics. Beginning with Berry's discovery of geometric phases in Quantum Mechanics [37] and Hannay's classical formulation of geometric holonomy [38], the geometric concepts have expanded far beyond their original context. They now play an essential role in thermodynamic geometry [30, 32], geometric optimization of nonequilibrium protocols [34], nonlinear transport [33, 35], kinetic and optimal transport theory [31], geometric formulations of nonequilibrium dynamics [29], and numerous modern approaches to irreversible transport. These developments demonstrate that curvature, holonomy, and geometric structure frequently govern observable physical phenomena. Yet in virtually all existing formulations, the relevant geometric objects are introduced a priori, and the ensuing physical consequences are derived from an already established geometric framework.

These two major developments—the theory of finite memory and the geometry of transport—have therefore evolved along largely independent directions. On the one hand, memory theories explain how transport depends on its past without attributing any intrinsic geometric significance to that history. On the other hand, geometric transport theories investigate the physical consequences of curvature and holonomy without addressing how such geometric structures might themselves emerge from transport history. The possibility that finite memory could constitute the microscopic origin of transport geometry has therefore remained essentially unexplored. This observation naturally raises a fundamental question:

Can finite memory generate the geometric structure governing transport, rather than merely modifying transport within a pre-existing geometric framework?

To address this question, we adopt a fundamentally different viewpoint. Rather than prescribing deformation from instantaneous local kinematics, we reconstruct transport from finite causal histories. Within this formulation, memory is no longer introduced as a constitutive correction acting on an existing transport law. Instead, causal transport history becomes the mechanism through which the geometric structure governing transport emerges. Geometry is therefore not postulated a priori but reconstructed directly from the ordered accumulation of transport histories. This change of perspective establishes a conceptual distinction between two complementary aspects of finite-memory transport. The first is dynamical, describing how past states influence subsequent evolution, as developed in conventional non-Markovian theories. The second is geometric, describing how causal transport histories generate the structure within which transport itself evolves.

In this picture, irreversibility is no longer taken solely as a consequence of constitutive dissipation, stochastic fluctuations, or nonlinear response. Instead, it becomes a manifestation of an emergent transport geometry generated by finite memory. Beyond its specific mathematical formulation, the present framework suggests a broader conceptual shift in nonequilibrium physics. Rather than treating memory as a temporal correction acting within an existing geometric description, finite memory becomes a mechanism capable of generating the geometry of transport itself. Causal transport history is elevated from a passive record of previous evolution to an active structural principle governing the organization of irreversible transport.

The paper is organized as follows. Section 2 develops the causal reconstruction of finite-memory transport and introduces the corresponding transport formalism. Sections 3–6 establish the resulting geometric framework and derive its principal mathematical properties. Section 7 discusses the relationship of the present theory to existing transport formalisms, while Section 8 examines its broader physical implications. Sections 9 and 10 present the numerical investigation and analyze the emergence of universal geometric behavior over a broad range of memory times. Finally, Section 11 summarizes the principal conclusions and outlines promising directions for future developments.

## 2 Memory-Dependent Reconstruction of Transport

### 2.1 Causal Reconstruction

In classical continuum mechanics, deformation is generated through local kinematic quantities evaluated at the present time. The velocity-gradient tensor is regarded as the infinitesimal generator of deformation and depends exclusively on the instantaneous state of motion. Within this framework, transport is intrinsically local in time. Memory effects, when incorporated, are introduced only at the constitutive level through stresses, fluxes, or transport coefficients, while the geometric description of deformation itself remains unchanged.

This local description ceases to be complete once transport possesses finite memory. If the present state depends on a finite interval of past evolution, the instantaneous velocity gradient can no longer provide a complete description of infinitesimal deformation. The generator of transport must instead be reconstructed from the causal accumulation of previous states.

To overcome this limitation, deformation is reconstructed from the ordered transport history experienced by each material element. The fundamental assumption is that the infinitesimal generator of transport depends on a finite memory horizon over which previous deformations contribute according to a causal weighting. This reconstruction is described by the memory-dependent transport connection

$$\mathrm{A_m}(\mathrm{x},\mathrm{t}) = \int_0^{\infty} \mathcal{K}(\tau)\nabla \mathrm{u}(\mathrm{x},\mathrm{t}-\tau)\mathrm{d}\tau \qquad (1)$$

where $\mathcal{K}(\tau)$ is a causal memory kernel satisfying $\mathcal{K}(\tau) = 0,\ \tau < 0$, together with the normalization condition, $\int_0^{\infty} \mathcal{K}(\tau)\ \mathrm{d}\tau = 1$.

The memory kernel determines how previous deformation states contribute to the reconstructed transport. Kernels concentrated near $\tau = 0$ recover short-memory behavior, whereas broader kernels progressively incorporate longer transport histories. No specific functional form is required at this stage. The subsequent geometric construction requires only causality together with the existence of a finite characteristic memory time.

Equation (1) should not be interpreted as a temporal averaging of the velocity gradient. Its role is conceptually different. The reconstructed quantity generates the infinitesimal evolution of transport after the complete causal history has been taken into account.

Finite memory therefore modifies not only the magnitude of deformation but also the mathematical object responsible for generating transport itself. This observation defines the conceptual departure from conventional non-Markovian formulations. Memory is not introduced as a correction acting on an already established kinematic description. Instead, finite memory defines the infinitesimal generator from which the geometry of transport is constructed. As a consequence, transport becomes intrinsically history-dependent before any constitutive assumptions are introduced.

Once deformation is reconstructed from causal transport history, temporal ordering becomes an intrinsic property of transport evolution rather than an external chronological parameter. As the following sections demonstrate, this ordered evolution provides the microscopic origin of noncommutativity, curvature, and geometric holonomy, thereby establishing the geometric foundation of irreversible transport developed throughout this work.

## 2.2 Memory as a Dynamical Connection

The reconstructed transport connection admits a natural geometric interpretation. Because it governs the evolution of infinitesimal deformations reconstructed from finite transport histories, $A_m$ may be interpreted as a dynamical connection defined on the space of transport histories, in the same mathematical sense that affine connections generate parallel transport in differential geometry [40].

Consider an infinitesimal material displacement $\delta x(t)$ transported along a trajectory.

Its evolution is governed by $\delta\dot{x}(t) = A_m(t)\,\delta x(t)$. The finite transport between two instants $(t_0, t)$ is therefore,

$$\mathcal{U}(t_0, t) = \mathcal{P} \exp\left(\int_{t_0}^{t} A_m(s)\, ds\right)$$

where $\mathcal{P}$ denotes chronological ordering. This expression reveals an essential consequence of finite memory. The resulting transport operator depends not only on the initial and final configurations but on the complete sequence of intermediate states connecting them.

The evolution is therefore determined by the history of transport rather than solely by its endpoints. Path dependence is thus not imposed as an additional hypothesis but emerges directly from causal reconstruction.

The memory-dependent connection therefore acquires the precise mathematical role of an affine connection. The memory kernel specifies how past deformation contributes locally to the connection, while chronological ordering determines how successive infinitesimal transport operations are composed. Geometry therefore emerges from the temporal organization of transport history itself. The classical local description is recovered continuously in the Markovian limit, $\mathcal{K}(\tau) \to \delta(\tau)$, for which $A_m(t) \to \nabla u(t)$.

This construction extends classical continuum kinematics while preserving its local limit.

The geometric significance of the connection becomes evident when transport generators evaluated at different times fail to commute. In this situation the ordered exponential cannot be reduced to the exponential of a single instantaneous generator. Instead, finite transport depends on the chronological composition of infinitesimal deformations, indicating that the evolution possesses an intrinsic geometric structure.

This observation establishes the geometric foundation of the theory developed throughout the remainder of the paper. Once transport is generated by memory-dependent transport connection, noncommutativity follows naturally from chronological composition, curvature emerges from the failure of successive transport operations to commute, and holonomy becomes the finite geometric signature of cyclic evolution. Within this framework, irreversible transport is no longer introduced through phenomenological constitutive assumptions but arises directly from the geometry generated by finite memory.

## 3 Geometry of Memory Transport

### 3.1 Ordered Transport and the Origin of Noncommutativity

Once transport is reconstructed from finite causal histories, chronological ordering ceases to be a technical feature of the evolution and becomes an intrinsic geometric property of transport itself. Because the memory-dependent transport connection evolves continuously with the accumulated transport history, finite evolution can no longer be generated by a single instantaneous operator. Instead, it is obtained through the ordered composition of history-dependent infinitesimal transport operations. The resulting finite evolution over a time interval $[0, T]$ is therefore described by

$$U(T) = \mathcal{P} \exp\left(\int_0^T A_m(t)\, dt\right)$$

The appearance of the ordering operator is not a mathematical convenience but a direct consequence of causal reconstruction. Since the transport connection continuously evolves as additional transport history is incorporated, the outcome of finite transport generally depends on the sequence in which infinitesimal deformations are applied. Chronological ordering therefore becomes an intrinsic component of transport evolution rather than an auxiliary mathematical prescription. This dependence disappears only if transport connections evaluated at different times commute,

$$[A_m(t_1), A_m(t_2)] = 0$$

In that case, the ordered exponential reduces to the exponential of a single effective generator, and the finite evolution depends only on the initial and final configurations, independently of the intermediate sequence of transport events. Finite memory, however, generally prevents this simplification because the reconstructed connection continuously changes as the causal transport history evolves. Hence, the transport operator is cast as an ordered exponential, with its underlying logarithmic generator expanded via the Magnus expansion,

$$U(T) = \exp\left(\sum_{i=1}^{\infty} \Omega_i\right)$$

whose first two contributions are,

$$\Omega_1 = \int_0^T A_m(t)\, dt, \qquad \Omega_2 = \frac{1}{2}\int_0^T dt_1 \int_0^{t_1} dt_2\, [A_m(t_1), A_m(t_2)]$$

The leading contribution ($\Omega_1$) represents the cumulative transport generated by the reconstructed connection, whereas, the second Magnus term ($\Omega_2$) has an entirely different origin. It arises exclusively from the failure of successive infinitesimal transport operations to commute and therefore constitutes the first correction that cannot be obtained from any instantaneous transport generator.

It measures the geometric contribution produced solely by chronological ordering.

An immediate consequence follows in the Markovian limit. When the transport connection becomes local in time, successive infinitesimal generators coincide and their commutator vanishes identically. The second Magnus contribution therefore disappears, $\Omega_2 \to 0$, recovering the classical description in which finite transport is completely determined by a single local generator. Noncommutativity therefore constitutes the first geometric manifestation of finite memory. It is not introduced through constitutive assumptions or phenomenological corrections but emerges inevitably from the causal reconstruction of transport itself. Once successive transport operations fail to commute, the emergence of curvature and geometric holonomy becomes a mathematical consequence of the ordered evolution, providing the foundation for the geometric theory developed in the following sections.

### 3.2 Memory-Induced Curvature

The commutator appearing in the second Magnus contribution admits a direct geometric interpretation. In differential geometry, the curvature of a connection measures the failure of successive infinitesimal parallel transport operations to commute and therefore characterizes the intrinsic non-integrability of the underlying connection [40]. The same mathematical structure emerges naturally in the present framework. The essential difference is that the connection is no longer prescribed on a pre-existing manifold but reconstructed directly from finite transport histories. This observation motivates the introduction of the memory-induced curvature operator,

$$\mathcal{R}(\mathrm{t}_1, \mathrm{t}_2) = [\mathrm{A}_\mathrm{m}(\mathrm{t}_1), \mathrm{A}_\mathrm{m}(\mathrm{t}_2)]$$

The curvature operator quantifies the extent to which finite transport cannot be represented as the accumulation of mutually commuting infinitesimal deformations.

Whenever $\mathcal{R}(\mathrm{t}_1, \mathrm{t}_2) = 0$, the reconstructed evolution becomes integrable in transport-history space, chronological ordering loses its geometric significance, and finite transport depends only on the initial and final configurations. In this limit, transport becomes effectively path-independent despite the existence of intermediate evolution.

The physical origin of the present curvature differs fundamentally from that encountered in conventional geometric transport theories. It is neither associated with spatial curvature nor generated by fluid vorticity, nor does it arise from an externally prescribed geometric manifold. Instead, it is produced entirely by the causal reconstruction of finite transport histories.

Finite memory therefore acts as the mechanism through which a non-integrable transport connection is generated, and curvature becomes the first intrinsic geometric quantity emerging from causal transport itself.

The Markovian limit follows immediately. As the memory kernel becomes localized, $\mathcal{K}(\tau) \to \delta(\tau)$, the reconstructed connection continuously approaches the instantaneous velocity gradient, $A_m(t) \to \nabla u(t)$, successive transport generators become effectively identical, chronological ordering ceases to produce geometric corrections, and the curvature vanishes continuously,

$$\mathcal{R}(t_1, t_2) \to 0$$

Classical continuum kinematics therefore appears as the zero-curvature limit of the present geometric framework rather than as a separate theoretical description.

The magnitude of the curvature is governed by the competition between two characteristic timescales: the forcing timescale $\omega^{-1}$, and the characteristic memory time $\tau_m$.

Their dimensionless product $\omega\tau_m$ therefore controls the extent to which finite memory distinguishes successive stages of transport evolution. Dimensional analysis consequently yields the generic scaling $\mathcal{R} \sim \omega\tau_m$, indicating that memory-induced curvature increases as transport history becomes progressively more effective at resolving the temporal structure of the deformation process.

Memory-induced curvature therefore constitutes the first intrinsic geometric observable generated by finite memory. It measures the impossibility of reducing causal transport histories to an equivalent instantaneous description and provides the fundamental geometric mechanism from which the subsequent emergence of holonomy and irreversible transport follows.

### 3.3 Holonomy as the Geometric Origin of Irreversible Transport

The physical consequences of memory-induced curvature become observable when transport is evaluated over a closed forcing cycle. In geometric transport theories, finite curvature implies that parallel transport around a closed loop generally fails to restore the initial state. The resulting mismatch is quantified by the associated holonomy.

Within the present framework, the relevant loop is not embedded in physical space but in transport-history space generated by periodic forcing. Even when the instantaneous flow remains locally irrotational throughout the cycle, the reconstructed transport connection evolves continuously because each infinitesimal deformation incorporates a different portion of the causal transport history. Successive transport operations therefore accumulate according to their chronological ordering, producing a finite holonomy after one complete forcing period. To leading order, the accumulated holonomy is determined by the second Magnus contribution, $\Omega_2 = \frac{1}{2}\int_0^T dt_1 \int_0^{t_1} dt_2 \left(\mathcal{R}(t_1, t_2)\right)$ which represents the total memory-induced curvature enclosed by the transport cycle. The resulting finite transport cannot be eliminated by simply reversing the instantaneous velocity field because its origin is global rather than local. It is determined by the ordered accumulation of the entire transport history rather than by any individual deformation event. Irreversibility therefore emerges as a geometric property of the complete transport cycle. For periodically driven transport, the observable manifestation of this geometric holonomy is the finite Lagrangian displacement,

$$\Delta\gamma \propto \frac{(\omega\tau_m)^2}{1 + (\omega\tau_m)^2}$$

whose magnitude depends only on the dimensionless parameter $(\omega\tau_m)$.

Three universal transport regimes follow naturally from this scaling. For $\omega\tau_m \ll 1$, the reconstructed connection approaches the instantaneous limit, memory-induced curvature becomes negligible, and the accumulated holonomy tends continuously to zero. Around $\omega\tau_m \sim 1$, the forcing period becomes comparable to the characteristic memory time. The reconstruction then distinguishes successive stages of the transport cycle most efficiently, producing the largest geometric accumulation. For $\omega\tau_m \gg 1$, the memory window averages increasingly many forcing cycles, leading to saturation of the accumulated displacement.

The central implication of the present framework is therefore clear. Irreversible transport does not require fluid vorticity, constitutive nonlinearities, stochastic forcing, or explicit symmetry breaking. It emerges as the observable manifestation of the holonomy generated by memory-induced curvature. Finite memory causes transport histories to enclose a finite geometric area in transport-history space, and the associated holonomy provides the measurable signature of the geometry generated by causal reconstruction.

## 4 Universal Geometric Transport Regimes

The previous sections established that finite memory reconstructs transport through a history-dependent connection whose ordered evolution gives rise to an intrinsic geometric structure. Once this framework has been established, a different question naturally arises. How does the resulting geometry evolve as the relative importance of memory changes?

Because the transport connection is reconstructed over a finite memory horizon, its geometric properties are expected to depend on the competition between the characteristic forcing timescale and the extent over which past transport contributes to the reconstruction. The objective of this section is therefore not to establish the existence of memory-induced geometry—which has already been demonstrated—but to identify the general principles governing its evolution. As shown below, the entire geometric response is organized by a single dimensionless variable, $\omega\tau_m$, which compares the characteristic timescale of the external forcing with the intrinsic memory timescale. This scaling reveals that transport generated by finite memory follows a universal organization that is largely independent of the specific physical system or the detailed form of the underlying memory kernel.

### 4.1 Universal Geometric Response

The dependence of transport on the single variable ($x = \omega\tau_m$) has an important physical consequence. It implies that systems operating on entirely different temporal scales may nevertheless exhibit identical geometric behavior once their dynamics are expressed in terms of the same dimensionless parameter. The transport process is therefore governed by relative temporal organization rather than by absolute timescales. The accumulated displacement consequently admits the normalized representation

$$\Delta\gamma = \Delta\gamma_{max}\, F(x), \qquad \text{with } F(x) = \frac{x^2}{1+x^2}$$

This normalized response separates two complementary aspects of the transport process.

The amplitude $\Delta\gamma_{max}$ contains the system-dependent information associated with the specific forcing and material properties, whereas $F(x)$ characterizes the intrinsic geometric response. Once normalized, the accumulated transport depends only on the competition between forcing and memory and becomes independent of their individual magnitudes.

The significance of this result extends beyond the particular model considered here.

It indicates that finite-memory transport possesses a universal geometric sector that is controlled exclusively by the relative temporal structure of the reconstruction process. Distinct physical systems, despite differing substantially in their microscopic dynamics, collapse onto the same normalized response whenever they share the same value of (x).

The function $F(x)$ therefore provides more than a convenient scaling law. It defines the universal organization of memory-generated transport, from which the different transport regimes emerge naturally as successive manifestations of the same underlying geometric mechanism.

### 4.2 Weak-Memory Regime

When $\omega\tau_m \ll 1$, memory operator samples only a narrow portion of the previous evolution, making the reconstructed connection almost indistinguishable from its instantaneous limit. Expanding the universal response $F(x) = x^2 + \mathcal{O}(x^4)$ immediately yields $\Delta\gamma \propto (\omega\tau_m)^2$

In this regime the transport connection changes only weakly over the memory interval. Successive transport operations therefore become nearly indistinguishable, their commutators remain small, and only negligible curvature is generated. The corresponding holonomy vanishes continuously as the memory time approaches zero. This limit recovers the local kinematics of classical continuum mechanics. The geometry generated by memory disappears smoothly, demonstrating that the present theory extends, rather than replaces, the conventional instantaneous description.

### 4.3 Optimal Temporal Matching

The strongest geometric transport occurs when $\omega\tau_m \sim 1$. At this scale the forcing period and the memory time become comparable.

The transport connection simultaneously retains information from previous stages of the evolution while remaining sufficiently responsive to the current forcing. Successive transport operations are therefore distinguished most efficiently by the reconstruction process. This optimal temporal matching maximizes the noncommutativity of the reconstructed connection, producing the largest curvature and, consequently, the greatest accumulated holonomy.

The resulting transport is therefore not enhanced by resonance in the conventional dynamical sense but by the most effective geometric discrimination of successive transport histories. This intermediate regime represents the point at which instantaneous dynamics and historical accumulation contribute with comparable weight, yielding the maximum efficiency of memory-induced geometric transport.

### 4.4 Long-Memory Regime

For $\omega\tau_{m} \gg 1$, the memory horizon extends over many forcing cycles. The reconstructed connection therefore incorporates an increasingly long transport history, causing additional temporal information to become progressively less distinguishable. Accordingly, $F(x) \rightarrow 1$ and the accumulated displacement approaches the finite limit $\Delta\gamma \rightarrow \Delta\gamma_{max}$

This saturation does not indicate unlimited growth of curvature. Rather, it reflects the finite resolving capability of the reconstructed connection. Once the memory horizon encompasses many forcing periods, additional history contributes little new geometric information, and further accumulation of holonomy becomes progressively inefficient.

The asymptotic regime therefore corresponds to a geometric saturation rather than to the disappearance of memory effects. Curvature remains finite, but its ability to generate additional irreversible transport is bounded by the finite temporal resolution of the transport reconstruction. Together, the weak-memory, optimal temporal matching and long-memory limits establish a universal classification of geometric transport generated by finite memory. Their common description through the single parameter $(x = \omega\tau_{m})$ demonstrates that the emergence of irreversible transport is governed not by the absolute magnitude of memory or forcing separately, but by their relative temporal organization.

## 5 Energetics and Intrinsic Geometry of Memory Transport

The preceding sections established that finite memory generates a transport connection whose ordered evolution gives rise to noncommutativity, curvature, and holonomy. These geometric structures provide a complete kinematic description of memory-induced transport.

A fundamental question nevertheless remains: does this geometry also possess an intrinsic energetic significance?

The present section addresses this question. Finite memory does not introduce an additional source of mechanical energy. Rather, it reorganizes the energetic structure of transport by modifying how deformation histories accumulate through the reconstructed connection.

The energetic contribution associated with finite memory is therefore not an independent constitutive effect but a direct manifestation of the geometry generated by causal transport reconstruction. The discussion below consequently extends the geometric framework developed thus far from a purely kinematic description to its intrinsic energetic interpretation.

### 5.1 Energetic Consequences of Memory Curvature

Consider an infinitesimal material displacement transported according to

$$\dot{\delta x}(t) = A_m(t)\, \delta x(t)$$

$A_m(t)$ is the memory-dependent transport connection introduced in Section 2. A natural quadratic transport measure is $E(t) = \frac{1}{2}\, \delta x^T(t)\, \delta x(t)$ whose time derivative becomes

$$\dot{E}(t) = \delta x^T\, A_m(t)\, \delta x$$

The accumulated transport activity during one forcing cycle is therefore

$$E_m = \int_0^T (\delta x^T A_m(t)\, \delta x)\, dt$$

This quantity reduces continuously to the classical energetic contribution when $K(\tau) \rightarrow \delta(\tau)$. Finite memory, however, fundamentally modifies the geometric organization of transport because the reconstructed connection generally does not commute at different times.

Consequently, the total transport energy naturally separates into, $E(T) = E_0(T) + E_R(T) + \cdots$ where the leading geometric contribution satisfies,

$$E_R \propto \delta x^T \Omega_2 \, \delta x$$

Substituting the curvature representation yields,

$$E_R \propto \frac{1}{2} \, \delta x^T \left( \int_0^T dt_1 \int_0^{t_1} dt_2 \, R(t_1, t_2) \right) \delta x$$

This result has an important conceptual implication. The additional transport energy is generated neither by instantaneous local deformation nor by phenomenological constitutive assumptions. It originates entirely from the curvature created by causal reconstruction of transport history. The energetic signature of memory is therefore a geometric consequence of the reconstructed connection.

### 5.2 Geometric Memory Action

Although $E_R$ characterizes the energetic effect of memory, it still depends on the particular displacement used to probe the transport. To isolate the intrinsic geometric contribution, we introduce a scalar functional that depends exclusively on the accumulated curvature.

The local geometric energy density is defined by $\varepsilon_{\mathcal{R}} = \mathrm{Tr}\,(\mathcal{R}^T \mathcal{R})$, which is positive by construction and vanishes only when the memory curvature itself vanishes. The accumulated geometric contribution is then measured through the memory geometric action

$$S_m = \frac{1}{2} \int_0^T dt_1 \int_0^{t_1} dt_2 \; \varepsilon_{\mathcal{R}}$$

Unlike the transport energy, $S_m$ depends only on the geometry generated by memory and is independent of any particular displacement field.

For monochromatic forcing, $A_m(t) = A \cos \omega t + B \sin \omega t$, the curvature becomes

$$\mathcal{R}(t_1, t_2) = [A, B] \sin \omega (t_1 - t_2)$$

leading to

$$S_m \propto \mathrm{Tr}\big([A,B]^T[A,B]\big)\,\frac{(\omega\tau_m)^2}{1+\,(\omega\tau_m)^2}$$

Introducing, $\varepsilon_{[A,B]} = \mathrm{Tr}([A,B]^T[A,B])$, the result assumes the compact universal form

$$S_m \propto \varepsilon_{[A,B]}\,\frac{(\omega\tau_m)^2}{1+\,(\omega\tau_m)^2}$$

This equation establishes a direct connection between the intrinsic noncommutativity of the reconstructed connection and the universal scaling law derived in Section 4. Geometry determines the tensorial structure, whereas finite memory governs its dynamical activation through the universal response function.

### 5.3 Representation-Independent Curvature Measures

The geometric action naturally introduces the positive scalar $\varepsilon_{\mathcal{R}} = \mathrm{Tr}\,(\mathcal{R}^T\mathcal{R})$, which measures the magnitude of memory curvature independently of the chosen orthogonal transport coordinates. Under the orthogonal transformation

$$A'_m = P^{-1}A_mP, \qquad P^T = P^{-1}$$

the curvature transforms as

$$\mathcal{R}' = [A'_m(t_1), A'_m(t_2)] = P^{-1}\,[A_m(t_1), A_m(t_2)]\,P = P^{-1}\mathcal{R}P$$

Using the cyclic invariance of the trace immediately gives

$$\varepsilon'_{\mathcal{R}} = \mathrm{Tr}\,(P^{-1}\mathcal{R}^T\mathcal{R}P) = \mathrm{Tr}\,(PP^{-1}\mathcal{R}^T\mathcal{R}) = \mathrm{Tr}\,(\mathcal{R}^T\mathcal{R}) = \varepsilon_{\mathcal{R}}$$

Therefore, $\varepsilon_{\mathcal{R}} = \mathrm{Tr}\,(\mathcal{R}^T\mathcal{R})$ constitutes a positive representation-independent measure of the magnitude of memory curvature. Unlike the observable transport displacement, this quantity characterizes the geometry itself. It quantifies the strength of the noncommutativity encoded in reconstructed transport histories and vanishes if and only if the curvature vanishes.

### 5.4 Intrinsic Geometric Invariant

An even stronger invariant can be constructed by removing not only coordinate dependence but the dependence on the entire linear representation of transport.

Define $I_m = \mathrm{Tr}\,(\mathcal{R}^2)$, and for an arbitrary invertible transformation $P \in GL(n)$, under which $A'_m = P^{-1} A_m P$, the curvature transforms covariantly according to $\mathcal{R}' = P^{-1}\mathcal{R}P$, so that

$$\mathcal{R}'^2 = (P^{-1}\mathcal{R}P)(P^{-1}\mathcal{R}P) = P^{-1}\mathcal{R}^2 P$$

The resulting invariant takes the form $I'_m = \mathrm{Tr}\,(\mathcal{R}'^2)$, which, upon invoking the cyclic property of the trace, directly yields,

$$I'_m = \mathrm{Tr}(\mathcal{R}'^2) = \mathrm{Tr}(P^{-1}\mathcal{R}^2 P) = \mathrm{Tr}(\mathcal{R}^2 P P^{-1}) = \mathrm{Tr}(\mathcal{R}^2) = I_m$$

Hence, $I_m = \mathrm{Tr}(\mathcal{R}^2)$ is invariant under the full general linear group $GL(n)$. It therefore characterizes an intrinsic geometric property of memory curvature rather than a particular representation of the transport dynamics. The distinction between the geometric quantities introduced in this section is now clear. The displacement $\Delta\gamma$ describes the observable manifestation of memory-induced transport. The scalar $\varepsilon_{\mathcal{R}}$ measures the positive magnitude of the underlying curvature. At the most fundamental level, $I_m$characterizes the intrinsic geometry of that curvature independently of any coordinate system or transport representation. The relation

$$S_m \propto \varepsilon_{\mathcal{R}} \frac{(\omega\tau_m)^2}{1 + (\omega\tau_m)^2}$$

demonstrates that the energetic activation of memory geometry follows exactly the same universal response function governing the observable displacement derived previously [42]. Geometry and transport therefore obey a common scaling law, while remaining distinguished by the intrinsic curvature encoded in $I_m$. The complete geometric hierarchy developed throughout this work can therefore be summarized as

$$A_m \rightarrow \mathcal{R} \rightarrow \varepsilon_{\mathcal{R}} \rightarrow S_m \rightarrow I_m \rightarrow \Delta\gamma$$

Finite memory therefore generates not only irreversible transport but also an intrinsic geometric structure endowed with its own energetic interpretation, scalar measures, and representation-independent invariants.

## 6 Universal Scaling Law for the Intrinsic Curvature Invariant

The previous section established that the intrinsic geometry generated by finite memory is characterized by the invariant $I_m = \mathrm{Tr}(\mathcal{R}^2)$, which is independent of the representation chosen for the transport dynamics. We now derive an explicit analytical expression for this invariant and show that its dependence on finite memory is governed by a universal scaling law.

To render the problem analytically tractable, we adopt a minimal periodic transport field $\nabla u(t) = A\cos\omega t + B\sin\omega t$, where $(A, B)$ are static transport generators, and $\omega$ is the fundamental driving frequency. This harmonic representation constitutes the simplest model capable of generating noncommuting transport histories while retaining full analytical tractability. For the exponential memory kernel $\mathcal{K}(\tau) = \frac{1}{\tau_m}\exp(-\tau/\tau_m)$, the reconstructed transport connection becomes

$$A_m(t) = \frac{1}{1 + (\omega\tau_m)^2}\,[(A + \omega\tau_m B)\cos\omega t + (B - \omega\tau_m A)\sin\omega t]$$

Substituting this expression into the definition of the memory curvature immediately yields

$$\mathcal{R}(t_1, t_2) = \frac{\omega\tau_m}{1 + (\omega\tau_m)^2}\,[A, B]\sin(\omega(t_1 - t_2))$$

This explicit solution confirms the general geometric arguments developed in the previous sections. All dependence on finite memory is contained in the dimensionless parameter $\omega\tau_m$**,** whereas the tensorial structure is determined entirely by the commutator of the transport generators. Using the intrinsic invariant introduced in Section 5, one obtains

$$I_m(t_1, t_2) = \mathrm{Tr}([A, B]^2)\left(\frac{\omega\tau_m}{1 + (\omega\tau_m)^2}\right)^2 \sin^2\big(\omega(t_1 - t_2)\big)$$

The system-specific geometric information is entirely contained in the constant

$I_0 = \mathrm{Tr}([A, B]^2)$, allowing the invariant to be written as

$$I_m(t_1, t_2) = I_0\left(\frac{\omega\tau_m}{1 + (\omega\tau_m)^2}\right)^2 \sin^2\big(\omega(t_1 - t_2)\big)$$

Averaging over one complete forcing cycle gives

$$I_m = I_0 \times \frac{1}{2}\left(\frac{\omega\tau_m}{1 + (\omega\tau_m)^2}\right)^2$$

This equation constitutes the universal scaling law for the intrinsic curvature invariant.

It naturally separates the invariant into two complementary contributions. The constant $I_0$ contains the system-dependent noncommutative structure of the transport generators (intrinsic geometry), whereas the dimensionless factor describes the universal activation of this geometry by finite memory.

The scaling law predicts a non-monotonic dependence of the invariant on the memory strength. The invariant vanishes in the instantaneous limit, increases as finite memory progressively distinguishes successive transport histories, reaches its maximum under optimal temporal matching $\omega\tau_m \sim 1$, and decreases again when the memory horizon becomes sufficiently long that additional transport history contributes progressively less new geometric information. This analytical result establishes the universal reference for the intrinsic geometry of memory transport. In the following numerical section, it provides the theoretical reference against which the emergence, maximum, saturation, and susceptibility of the curvature invariant are quantitatively examined. The analytical derivation presented above is based on a monochromatic forcing as the minimal exactly solvable realization of the memory-induced transport geometry. This assumption is introduced solely to obtain closed-form expressions for the reconstructed connection and the associated curvature invariant.

The analytical solution is specific to monochromatic forcing, whereas the geometric framework itself is considerably more general. For an arbitrary time-dependent velocity-gradient field, the memory connection remains defined through the causal reconstruction $A_m(t)$, and the curvature is universally given by $\mathcal{R}(t_1, t_2)$.

In this more general setting, different temporal modes contribute simultaneously to the reconstructed connection, and the resulting curvature naturally becomes a superposition of commutators between dynamically coupled transport modes. The intrinsic invariant retains the universal definition $I_m = \mathrm{Tr}(\mathcal{R}^2)$, independently of the spectral complexity of the underlying flow.

The monochromatic solution derived here should therefore be viewed as the simplest analytically tractable realization of a broader geometric framework applicable to general non-harmonic transport fields.

## 7 Positioning the Present Framework

The preceding sections established that finite memory reconstructs the transport dynamics through a causal connection whose ordered evolution generates curvature, holonomy, energetic measures, and intrinsic geometric invariants. The resulting framework naturally raises the question of its relationship to existing descriptions of transport and irreversible dynamics. Rather than introducing an alternative constitutive description, the present framework proposes a different geometric foundation for transport. Its purpose is not to replace classical continuum mechanics or existing memory-based approaches, but to identify a geometric structure that remains implicit within them. In this perspective, finite memory is interpreted as the mechanism through which transport acquires an intrinsic differential geometry.

### 7.1 From Local Kinematics to Intrinsic Geometry

Classical continuum mechanics describe transport through the instantaneous velocity-gradient tensor evaluated at the current state of motion. Deformation is therefore regarded as a local kinematic quantity, while memory effects, when considered, enter only through constitutive relations governing stresses, fluxes, or material properties. The geometric structure of transport itself is assumed from the outset and remains independent of the temporal history of the flow. The present framework adopts a fundamentally different viewpoint. Instead of prescribing transport locally, it reconstructs the transport connection from a finite causal history $A_m(x, t)$. The geometry governing transport is therefore not assumed a priori but emerges dynamically through the ordered accumulation of past deformations.

The reconstructed connection subsequently generates noncommutativity, curvature, and holonomy as intrinsic consequences of causal transport reconstruction.

The classical description is recovered continuously in the limit $\tau_m \to 0$, for which the reconstructed connection reduces to the instantaneous velocity gradient and the memory curvature vanishes identically. The present framework therefore extends classical continuum kinematics while preserving its local limit.

### 7.2 Memory as a Generator of Geometry

Finite memory plays a central role in several existing theoretical frameworks, including viscoelastic continuum models, non-Markovian transport theories, and geometric-phase formulations. In viscoelastic theories, memory modifies the constitutive response of materials after the deformation has already been specified. Modern non-Markovian theories describe memory through a variety of complementary frameworks, including generalized Langevin equations, generalized master equations, and more recently process-tensor formulations capable of describing multi-time quantum memory effects [8–11, 15–28].

In geometric-phase theories, the connection responsible for holonomy is introduced as part of the underlying mathematical description, and geometric phases are subsequently derived from its properties. Together, these approaches deliver a refined description of memory-induced correlations and information backflow beyond the Markovian paradigm.

For all their conceptual differences, they converge on a single goal: deciphering how memory governs dynamical evolution within a given theoretical framework. The present framework reverses this logical sequence. Memory acts directly on the reconstruction of transport itself, allowing the connection to emerge from causal history before any constitutive assumptions are introduced. Curvature is therefore not imposed geometrically nor inherited from an external parameter space; it is generated internally by the noncommutative composition of memory-dependent transport histories [40-41]. Within this perspective, finite memory is elevated from a constitutive correction or statistical descriptor to the geometric mechanism responsible for the emergence of transport geometry.

### 7.3 Conceptual Position of the Present Theory

The conceptual architecture developed throughout this work may be summarized by the sequence:

$$\text{Memory} \rightarrow \mathrm{A_m} \rightarrow \mathcal{R} \rightarrow \mathrm{I_m} \rightarrow \Delta\gamma$$

This sequence should be interpreted as a logical hierarchy rather than a temporal evolution, identifying the successive geometric structures generated by causal transport reconstruction. Finite memory first reconstructs the transport connection. The ordered evolution of this connection generates noncommutative transport, giving rise to a memory-induced curvature.

This curvature possesses intrinsic scalar measures and representation-independent invariants, whose accumulated holonomy produces the observable irreversible transport.

Within this framework, irreversibility does not require vorticity, constitutive nonlinearities, stochastic forcing, or explicit symmetry breaking. Instead, it emerges as the geometric consequence of finite-memory reconstruction.

The present theory should therefore be viewed not as an alternative constitutive model for transport, but as a geometric extension of continuum kinematics. Existing continuum, viscoelastic, geometric-phase, and non-Markovian theories remain fully consistent with this perspective; the present framework complements them by identifying the intrinsic differential geometry generated by causal transport history.

This perspective provides a unified geometric interpretation of finite-memory transport, in which the connection, curvature, energetic measures, intrinsic invariants, and irreversible displacement all emerge from a single organizing principle: the causal reconstruction of transport through finite memory.

## 8 Discussion and Outlook

The present work has introduced a geometric formulation of finite-memory transport in which causal reconstruction transforms transport history into an intrinsic differential geometry. Rather than modifying constitutive laws or introducing phenomenological corrections, finite memory reconstructs the transport connection itself, from which curvature, holonomy, energetic measures, and intrinsic invariants emerge naturally. This perspective suggests that, whenever finite memory is present, irreversibility may be interpreted as a geometric consequence of transport history rather than solely as a consequence of constitutive complexity or dissipative mechanisms [12].

### 8.1 From Transport Theory to Transport Geometry

Classical continuum transport theories describe deformation through local kinematic quantities defined instantaneously in time. In contrast, the present framework reconstructs transport from finite causal histories, reinterpreting the velocity-gradient field to a history-dependent connection. Geometry is therefore no longer prescribed at the level of the underlying continuum description but emerges dynamically through the ordered composition of memory-dependent transport. This shift in perspective changes the role of memory fundamentally. Instead of acting as a correction to an already established transport process, memory becomes the mechanism through which the geometric structure of transport itself is generated. Connection, curvature, holonomy, energetic measures, and intrinsic invariants all arise from this single organizing principle. The resulting framework therefore extends classical continuum kinematics into a geometric description of transport history.

### 8.2 Towards a Geometry of Memory-Controlled Dynamics

Throughout the present work, the characteristic memory time has been treated as the parameter governing the emergence of curvature and irreversible transport. In realistic physical systems, however, the memory time is itself determined by underlying material relaxation processes and therefore depends on thermodynamic variables,

$$\tau_m = \tau_m(T, P, \eta, \varrho, \dots)$$

The specific functional dependence is expected to be system dependent and is not addressed in the present work. Consequently, the reconstructed transport connection, its curvature, and the associated geometric invariants inherit an implicit thermodynamic dependence,

$$A_m, \mathcal{R}, I_m = A_m, \mathcal{R}, I_m(T, P, \rho, \eta)$$

This observation suggests that thermodynamic variables may regulate not only transport coefficients but the geometry of transport itself. Developing a thermodynamic theory of memory-generated geometry therefore represents a natural continuation of the present framework. An equally important consequence concerns the governing equations of continuum dynamics. Since transport is reconstructed through the memory-dependent connection, it is natural to ask whether the transport operators appearing in continuum mechanics should themselves be reconstructed from finite histories. Such an extension would promote memory from a constitutive ingredient to a structural component of the dynamical equations, opening the way of a geometric formulation of continuum transport with memory.

### 8.3 Broader Perspectives

The present work therefore provides a complementary geometric viewpoint on the long-standing problem of irreversible transport [12].

Although the analytical developments presented here have focused on periodically driven transport, the geometric construction relies only on two fundamental ingredients: **causal reconstruction** and **temporal ordering**.

The underlying geometric construction is therefore expected to remain applicable far beyond the specific examples analyzed in this work. Potential directions include complex fluids, viscoelastic media, active matter, biological transport, turbulent flows, soft condensed matter, and other nonequilibrium systems in which finite memory influences dynamical evolution. More generally, the present approach suggests that memory may represent a universal geometric mechanism through which transport acquires an intrinsic geometric structure.

Beyond these immediate applications, the theory points toward a broader conceptual perspective. In most existing geometric formulations of physics, geometry is associated primarily with the structure of physical or parameter spaces. The present work suggests an alternative possibility: geometry may also emerge from the temporal organization of transport history itself. Within this viewpoint, finite memory is not merely a temporal correction acting on transport but the physical mechanism through which transport generates its own geometry.

If this perspective proves to extend beyond the examples considered here, it could provide the basis for a unified geometric description of irreversible transport in which causal memory, rather than constitutive complexity alone, becomes the fundamental origin of connection, curvature, holonomy, and intrinsic geometric invariants.

## 9 Numerical Results and Scaling Analysis

The theoretical framework developed in the preceding sections leads to four principal predictions: (i) the emergence of memory-induced transport curvature, (ii) the appearance of intrinsic geometric invariants, (iii) universal scaling governed by the dimensionless parameter ($\omega\tau_m$), and (iv) the existence of a long-memory saturation regime. The purpose of the present section is to examine these predictions through direct numerical simulations.

Because the present theory makes explicit quantitative predictions, numerical analysis provides a direct means of assessing both the existence and the robustness of the proposed geometric mechanisms. Rather than serving as independent numerical illustrations, the simulations provide a systematic assessment of the geometric framework developed throughout this work.

We commence by examining the emergence of curvature in transport-history space, then investigate its cumulative geometric effects, its universal scaling behavior, and finally the mechanisms responsible for the asymptotic saturation of memory-induced transport. Taken together, the numerical results demonstrate that the geometric structures derived analytically constitute observable and quantitatively robust consequences of finite-memory transport.

### 9.1 Emergence of Memory-Induced Geometry

The first objective of the numerical analysis is to determine whether the geometric structures predicted by the theoretical framework emerge from the reconstructed transport dynamics. The analysis therefore focuses on the curvature generated by the memory-dependent transport connection introduced in Sections 2 and 3, which constitutes the fundamental geometric signature of finite-memory transport.

Fig 1 presents the reconstructed curvature field $\mathcal{R}(t_1, t_2)$ for increasing values of the characteristic memory time $\tau_m$. The simulations reveal two systematic trends. First, the amplitude of the curvature increases continuously as the memory horizon is extended. Second, the region of transport-history space occupied by nonzero curvature expands progressively with increasing memory time. Together, these observations provide the first quantitative evidence that the geometric structure predicted by the theoretical framework emerges continuously as progressively longer transport histories are incorporated into the reconstruction. More importantly, the numerical results confirm the central prediction of the theory: finite memory does not merely rescale the magnitude of transport; instead, it generates an intrinsic geometric structure whose extent grows continuously as longer transport histories are incorporated into the reconstruction. The emergence of curvature thus constitutes the first experimentally accessible signature of the geometric framework developed in this work, providing the foundation for the global analyses presented in the following sections.

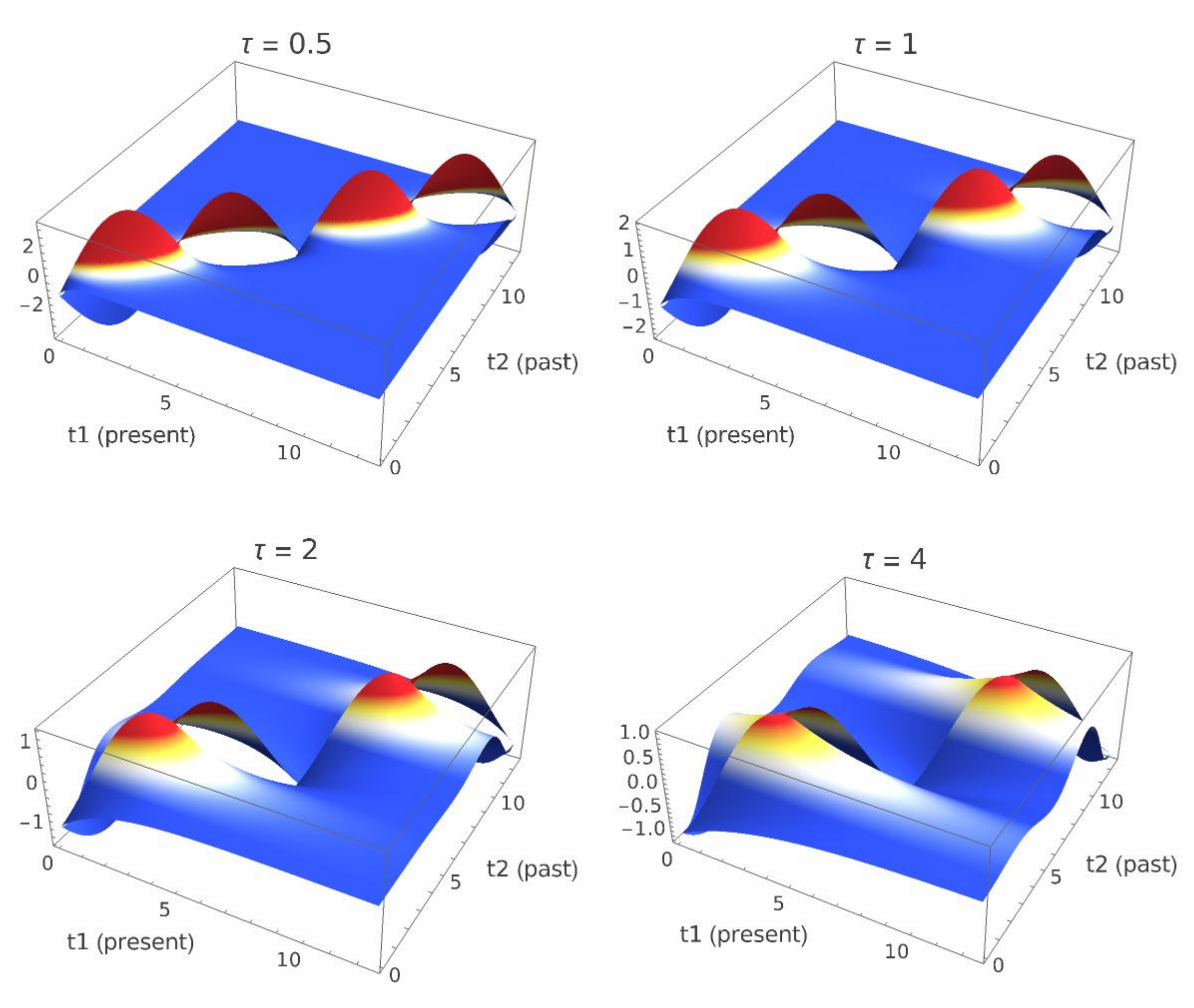

Fig 1- *Memory-induced curvature field* $\mathcal{R}(\mathbf{t_1}, \mathbf{t_2})$ *in transport-history space for increasing memory times: The progressive amplification and expansion of the curvature demonstrate the emergence of an intrinsic geometric structure generated by causal memory reconstruction*

Although the curvature field establishes the local emergence of memory-induced geometry, its amplitude alone does not provide an intrinsic measure of the underlying geometric structure.

A representation-independent characterization requires the curvature invariant introduced in Section 6. The corresponding invariant field is presented in Fig 2. Consistent with the analytical predictions, the invariant exhibits the same systematic enhancement observed in the curvature itself. Increasing memory strengthens not only the local curvature but also its global geometric measure, demonstrating that the numerical evolution faithfully reproduces the intrinsic geometric structure predicted by the theory.

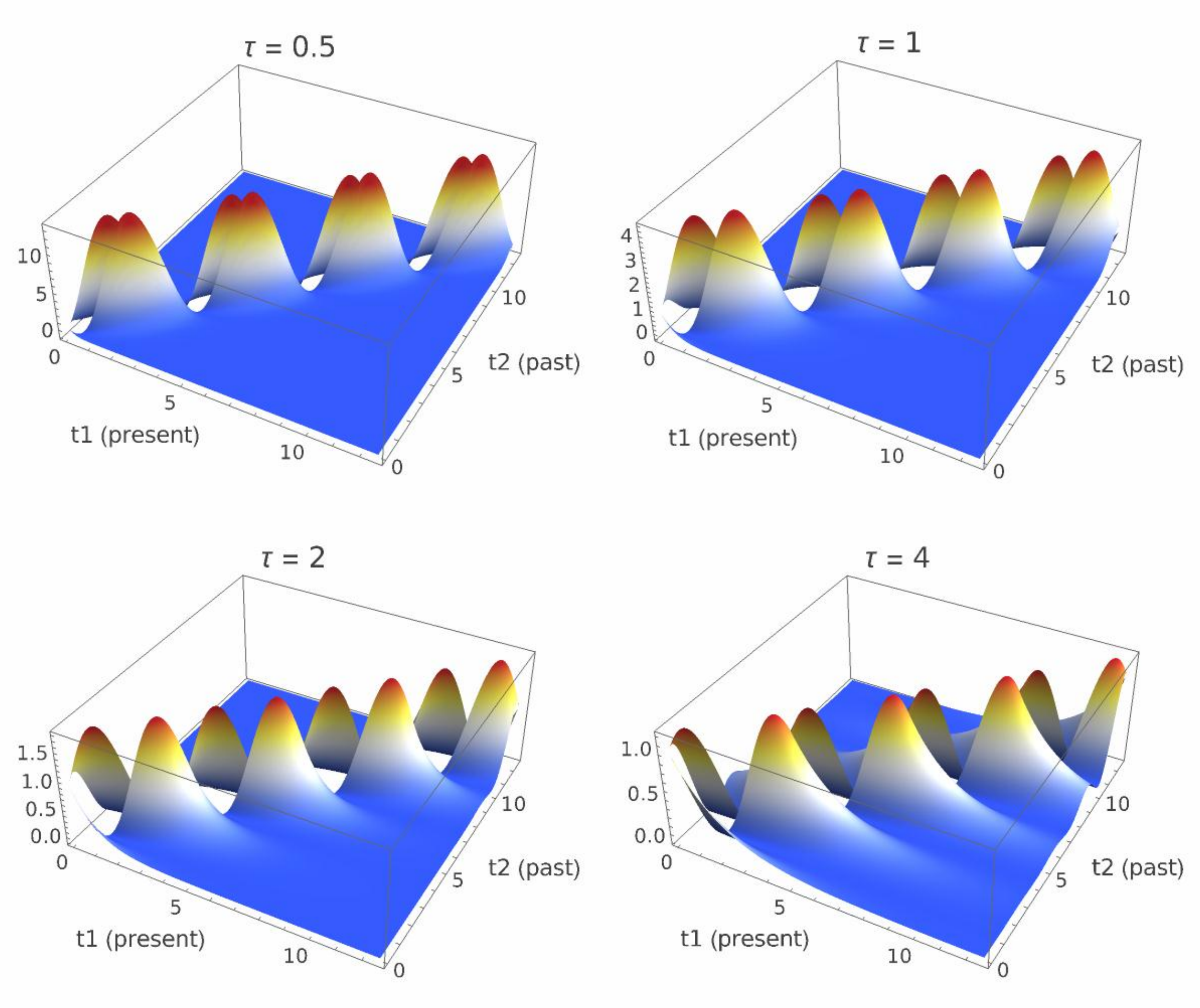

Fig 2- *Memory-induced geometric invariant* $\mathbf{I_m(t_1, t_2)}$ *associated with the curvature field. The amplification of the invariant confirms that finite memory produces a systematic increase in the intrinsic geometric deformation of transport history*

To isolate the specific contribution generated by memory, Fig 3 presents the difference between the finite-memory curvature and its memoryless counterpart ($\Delta\mathcal{R} = \mathcal{R}(\tau_m) - \mathcal{R}(\tau_{ref}), \tau_{ref} = 1$). The resulting distribution is dominated by positive values that progressively expand throughout the history manifold as the memory time increases.

This result demonstrates that memory enhances curvature through a cumulative geometric mechanism rather than through localized fluctuations or oscillatory corrections. Irreversible transport thus develops through the progressive accumulation of geometric contributions distributed throughout transport-history space.

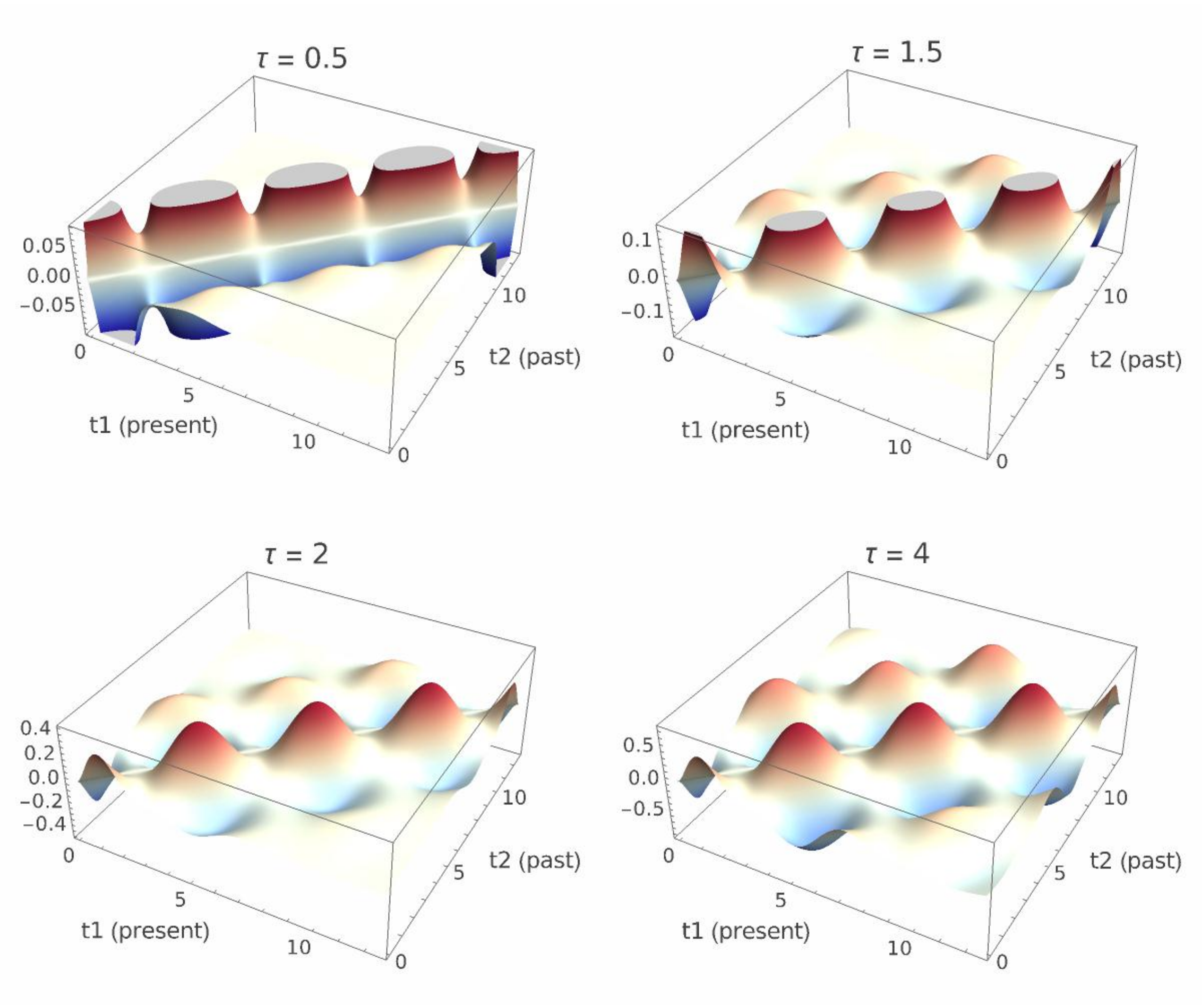

Fig 3- *Difference between the finite-memory curvature and the instantaneous (memoryless) curvature: The dominance of positive contributions demonstrates that finite memory acts constructively in generating transport curvature*

A complementary measure of this amplification is obtained from the memory sensitivity of the curvature ($S_{\mathcal{R}} = \frac{\partial \mathcal{R}}{\partial \tau_m}$) displayed in Fig 4. The sensitivity remains predominantly positive throughout the transport-history manifold, indicating that increasing memory systematically reinforces the reconstructed curvature. Equally important, the positive response extends over a large portion of transport-history space rather than remaining localized around isolated regions. Finite memory therefore acts as a control parameter governing the development of transport geometry rather than as a passive temporal correction to otherwise local dynamics.

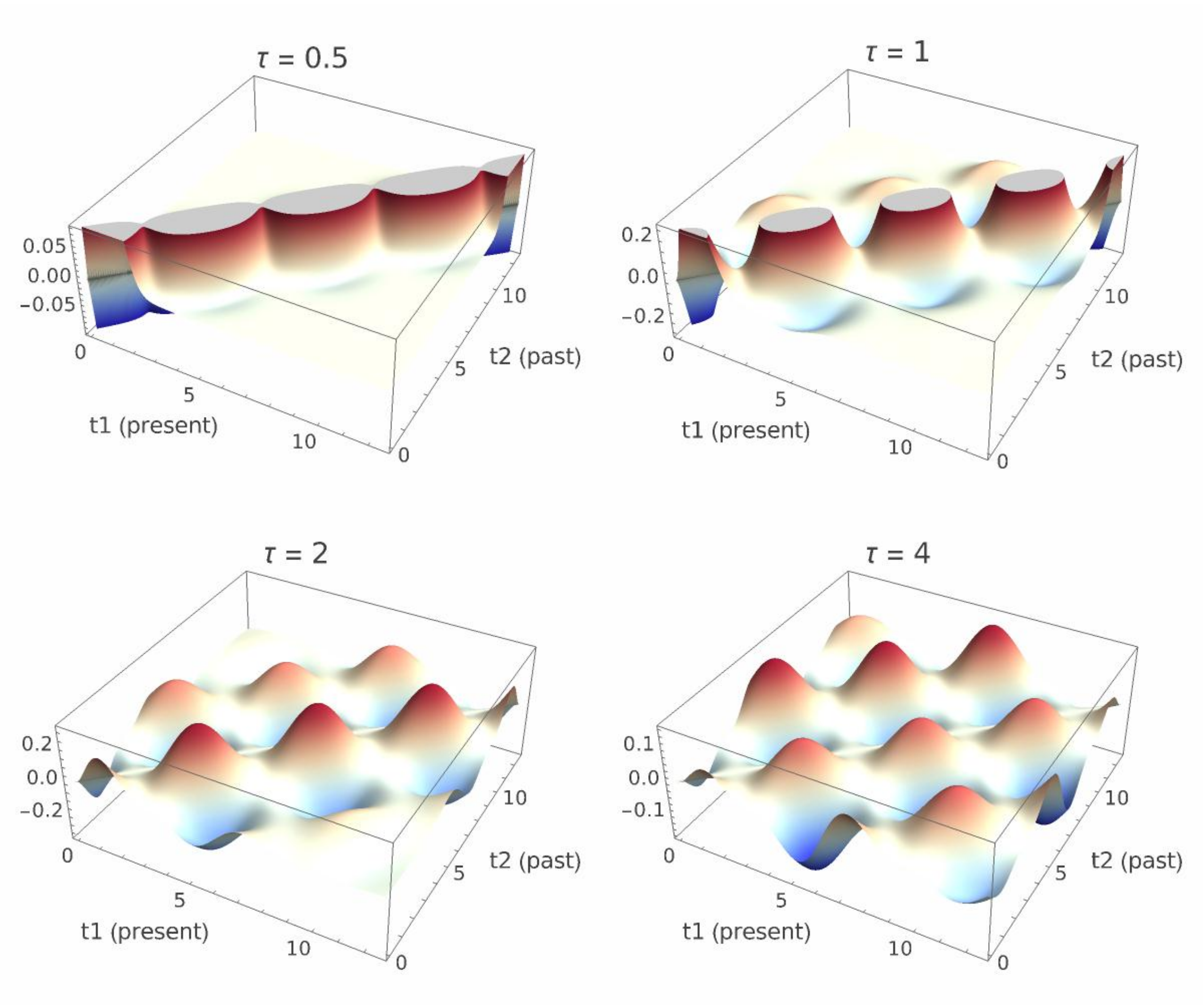

Fig 4-*Memory sensitivity of the reconstructed curvature: The predominantly positive response demonstrates that increasing memory systematically strengthens the geometric structure generated by transport history*

These Figs. 1–4 collectively establish a coherent sequence of numerical evidence supporting the geometric framework developed in the preceding sections. The reconstructed curvature first emerges locally within transport-history space, its intrinsic magnitude subsequently increases through the associated invariant, memory contributes constructively to its accumulation, and the positive sensitivity demonstrates that this evolution is systematically reinforced as the memory horizon increases. Collectively, these results provide the first quantitative validation of the central theoretical prediction that finite memory generates an intrinsic transport geometry through causal reconstruction.

### 9.2 Global Quantification of Memory-Induced Geometry

The local emergence of memory-induced curvature established in the previous subsection provides the first numerical evidence for the geometric framework developed in this work. The next question is whether this local structure also gives rise to measurable global geometric observables. While the curvature field provides direct evidence for the existence of this geometry, a quantitative characterization of its global evolution requires scalar observables that quantify the cumulative strength of the reconstructed transport over the entire computational domain. The first such observable is the spatially averaged geometric invariant, $< \mathrm{I_m} >$ whose dependence on the characteristic memory time is presented in Fig 5.

Unlike the local curvature maps discussed previously, this quantity provides a global measure of the intrinsic geometric deformation generated by memory reconstruction. The simulations reveal a continuous monotonic increase of the averaged invariant with increasing memory time. This behavior demonstrates that progressively incorporating longer transport histories systematically strengthens the overall geometric structure of the reconstructed dynamics. More significantly, the monotonic evolution of $< \mathrm{I_m} >$ indicates that memory does not merely redistribute curvature within transport-history space. Instead, it progressively increases the intrinsic geometric content of the reconstructed transport. The averaged invariant thus provides the first global quantitative measure of the cumulative geometric organization generated by finite memory.

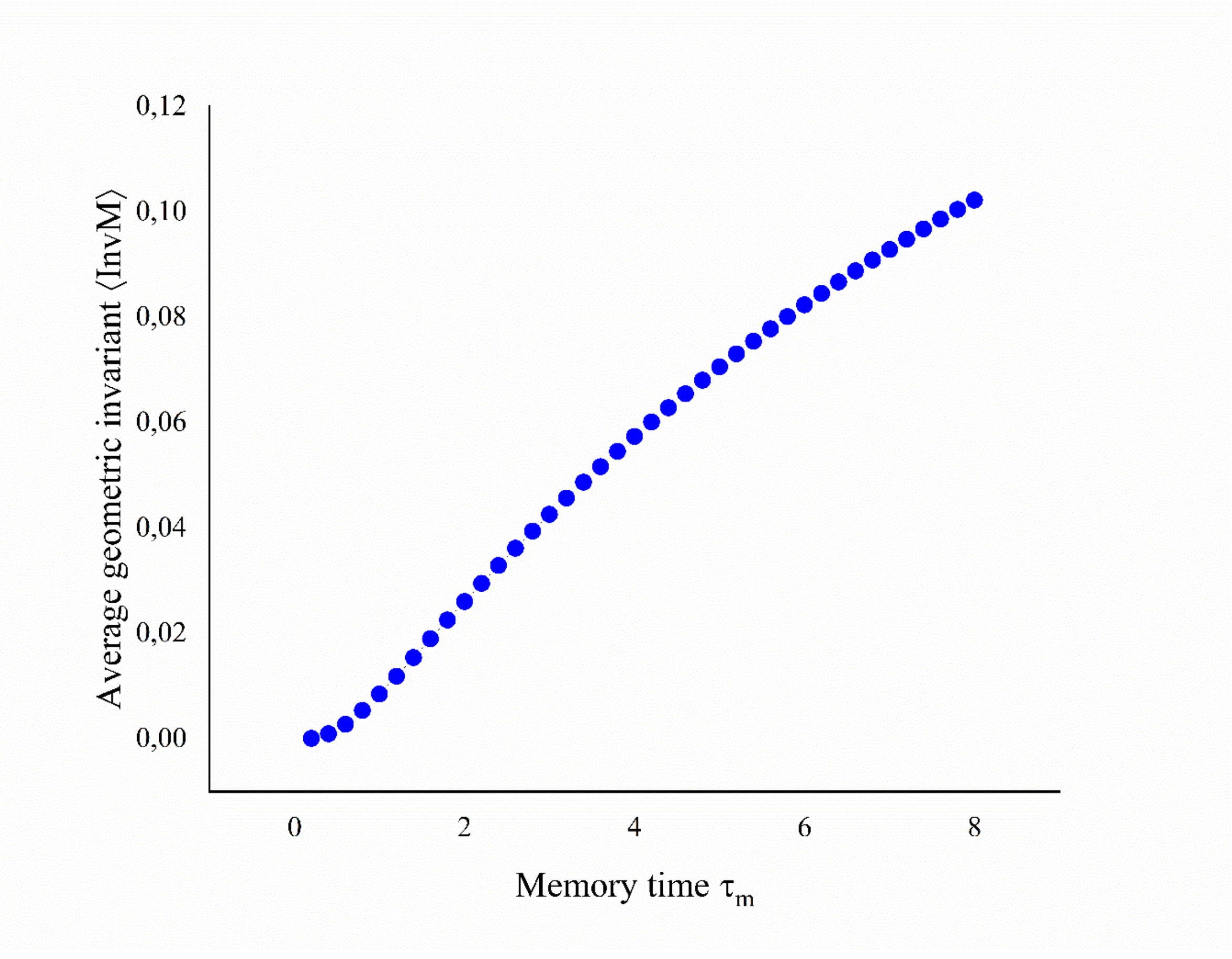

Fig 5- *Average geometric invariant* $< I_m >$ *as a function of the memory time* $\tau_m$*: The monotonic increase quantifies the progressive global accumulation of memory-induced transport geometry*

To further characterize this cumulative behavior, Fig 6 presents the integrated geometric invariant obtained by accumulating the numerical invariant over the transport history. Whereas the averaged invariant measures the instantaneous global strength of the reconstructed geometry, the integrated invariant quantifies the total geometric content generated during the evolution. The integrated invariant likewise increases monotonically with memory time, demonstrating that the geometric structures generated locally accumulate coherently throughout the transport history**.** No evidence of destructive interference or oscillatory cancellation is observed. Instead, successive transport histories reinforce one another, producing a progressive accumulation of irreversible geometric transport.

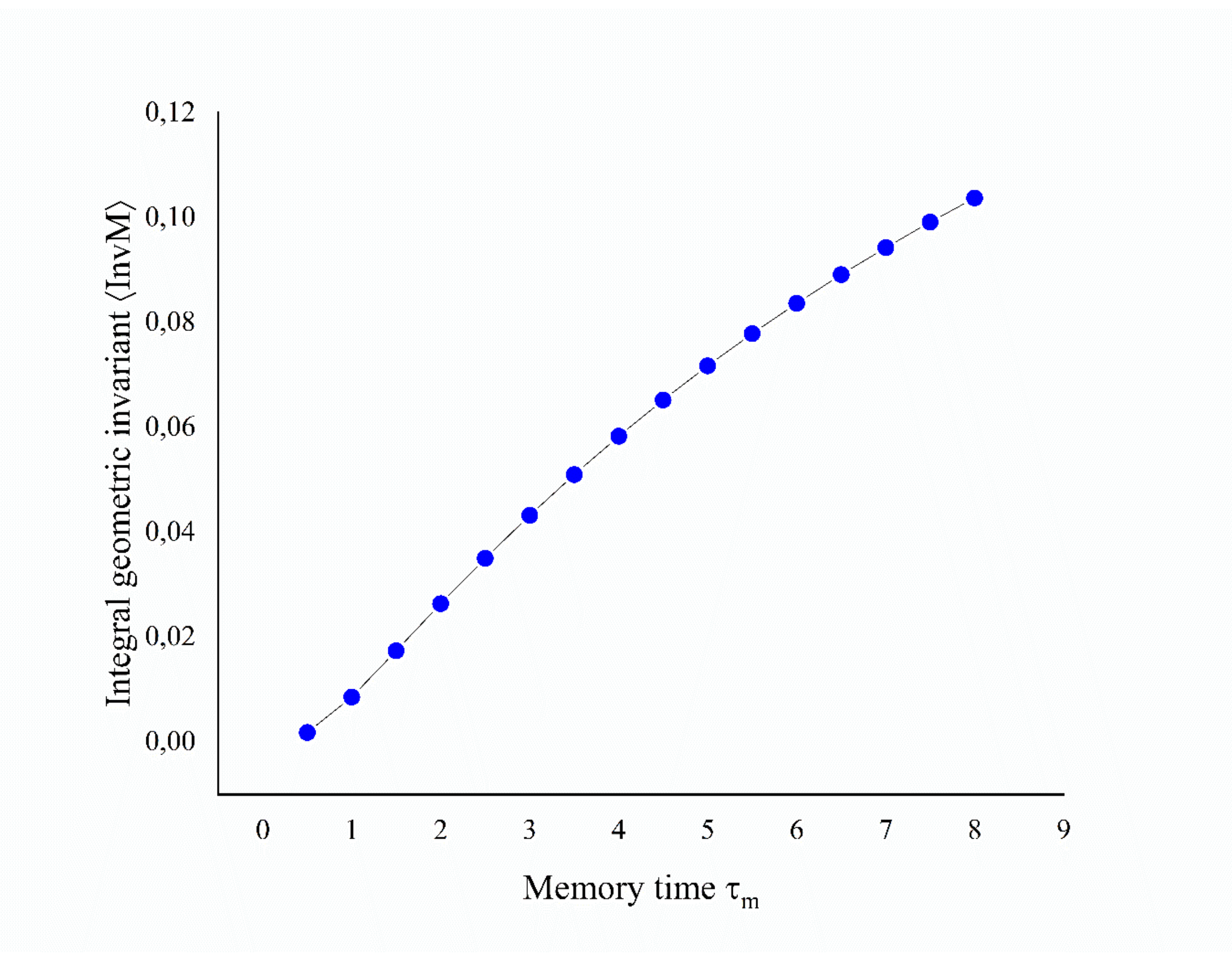

Fig 6- *Integrated geometric invariant as a function of the memory time. The continuous increase confirms that memory-induced curvature accumulates coherently throughout the transport history, leading to the coherent accumulation of global transport geometry*

The foregoing Fig 5 and 6 demonstrate that the local geometric structures identified in the previous subsection possess a well-defined global organization. The averaged invariant quantifies the instantaneous strength of the reconstructed geometry, whereas the integrated invariant measures its cumulative development over transport history. Their common monotonic evolution confirms that finite memory generates a coherent global geometric structure whose accumulation proceeds systematically as the memory horizon increases. These results therefore provide the first quantitative evidence that the geometric irreversibility predicted by the theory is not merely a local property but a robust global characteristic of memory-driven transport.

## 9.3 Universal Scaling and Emergent Saturation

Having established that the averaged and integrated geometric invariants increase monotonically with the memory time, we now address a more fundamental question.

Is the memory-induced geometric response controlled independently by the forcing frequency and the memory time, or does its evolution collapse onto a universal geometric law governed by a single dimensionless parameter?

The analytical theory developed in Section 6 predicts that the essential control parameter is not the memory time itself, but the dimensionless combination $x = \omega\tau_m$, which measures the competition between the forcing timescale $\omega^{-1}$ and the characteristic memory timescale $\tau_m$. If this prediction captures the essential geometry of the problem, numerical results obtained under different physical conditions should collapse onto a single master curve when expressed as a function of x. This prediction is verified in Fig 7, where the averaged geometric invariant is replotted as a function of the dimensionless memory parameter $(\omega\tau_m)$.

Although the simulations correspond to different values of the individual control parameters, all numerical data collapse onto a single master curve. The observed collapse demonstrates that the geometric response is controlled primarily by the combined parameter $(\omega\tau_m)$, rather than by the forcing frequency or the memory time independently. The emergence of this master curve provides one of the strongest numerical validations of the analytical framework developed in the preceding sections. It confirms that the universal scaling predicted analytically is not restricted to a particular transport configuration but represents a general property of memory-induced geometric transport. The numerical collapse therefore reveals that memory-induced geometry exhibits a universal organization controlled entirely by the relative competition between forcing and memory timescales.

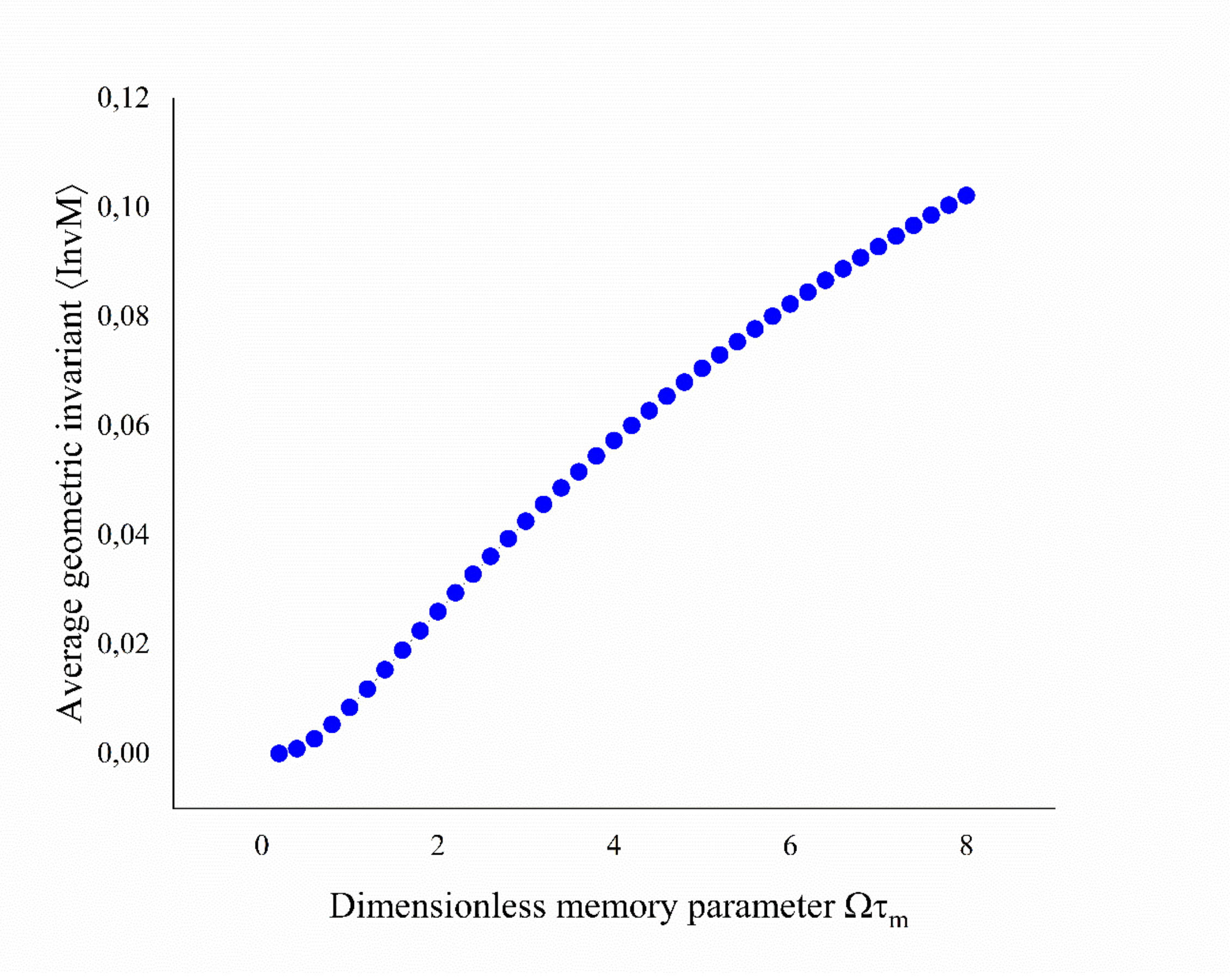

Fig 7-*Universal scaling collapse of the averaged geometric invariant as a function of the dimensionless memory parameter* $(\omega\tau_m)$*: The collapse of all numerical data onto a single master curve demonstrates that memory-induced geometric transport is governed primarily by the dimensionless combination* $(\omega\tau_m)$*, independently of the individual values of the physical parameters*

The universal scaling collapse demonstrated above establishes that memory-induced transport is governed by the single dimensionless parameter. This result reveals the existence of a universal geometric law, independent of the particular values of the forcing frequency and memory time. The collapse onto a universal master curve establishes the governing scaling variable but does not determine the ultimate evolution of the geometric response.

In particular, it remains unclear whether the reconstructed geometry continues to strengthen indefinitely or whether finite memory leads to an intrinsic asymptotic regime.

This question is addressed in Fig 8 by extending the numerical analysis far beyond the memory interval considered previously. Unlike the previous analysis, which established the existence of universal scaling, the present results explore the evolution of the geometric response deep into the long-memory regime.

The numerical data show that the averaged geometric invariant continues to increase as the memory window is extended. However, this increase is no longer characterized by the nearly uniform growth observed at shorter memory times. Instead, the rate of geometric amplification decreases progressively, indicating that the influence of additional transport history becomes gradually less effective. The reconstructed geometry therefore continues to evolve, but at a continuously diminishing rate. This evolution reveals an important geometric property of memory reconstruction. Finite memory does not produce an unlimited accumulation of geometric irreversibility. Rather, the reconstructed transport approaches an asymptotic geometric state in which successive extensions of the memory horizon generate progressively smaller geometric corrections. The numerical simulations therefore provide the first direct evidence that the universal geometric response possesses an intrinsic saturation regime.

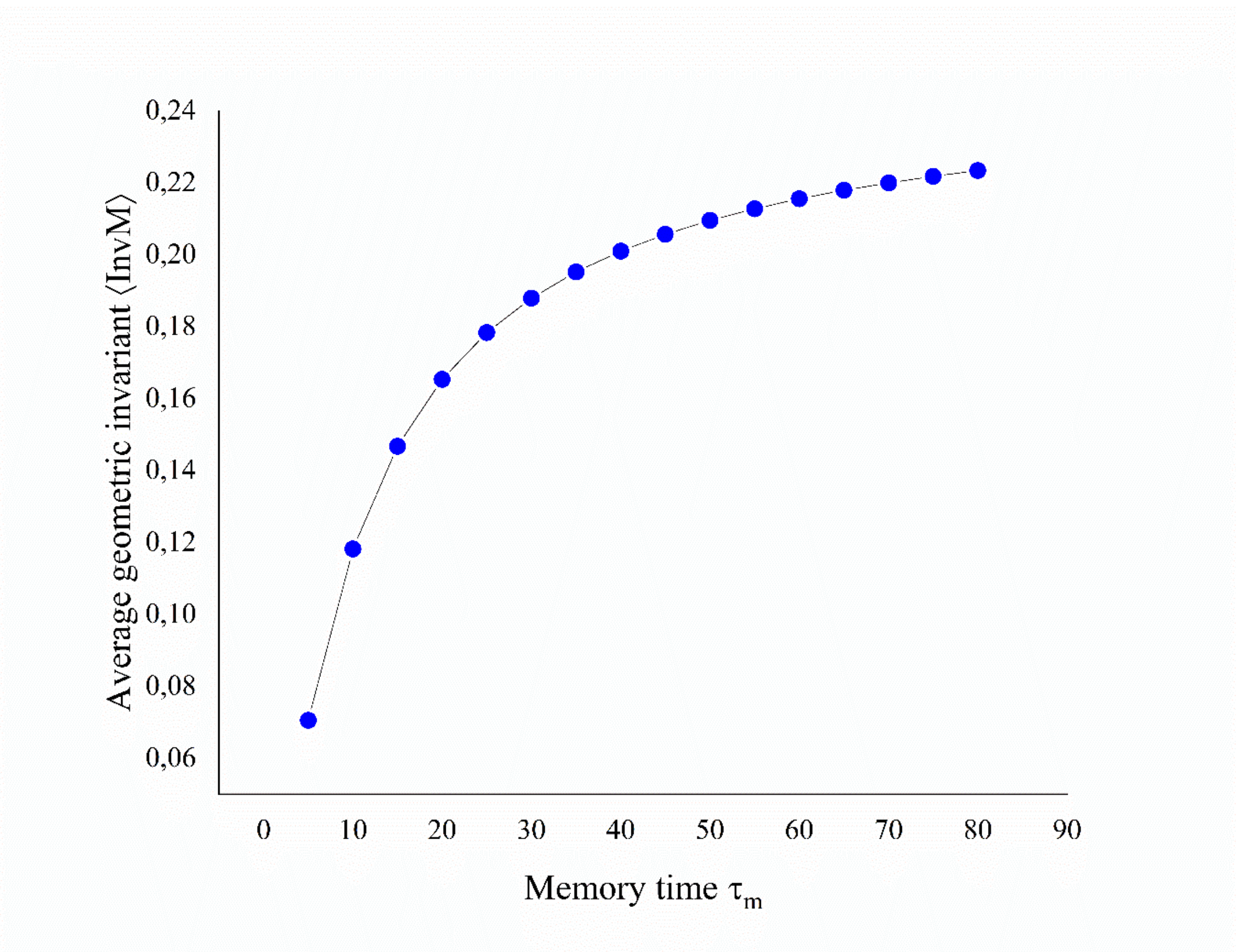

Fig 8- *Extended-memory behavior of the averaged geometric invariant over the interval* [0,80]*: Although the invariant continues to increase as progressively longer transport histories are incorporated, its growth gradually slows and approaches a finite asymptotic value, revealing the emergence of an intrinsic geometric saturation regime*

The existence of saturation naturally raises a further question: can this asymptotic behavior be described quantitatively by a simple geometric law?

Establishing such a law would not only characterize the approach toward saturation but would also identify the characteristic memory scale governing this transition. This issue is addressed in the following analysis through an explicit fitting of the numerical data. The nature of this saturation is examined quantitatively in Fig 9, where the numerical data are fitted by the exponential law:

$$< I_m > (\tau_m) = A\,(1 - \exp\left(-\frac{\tau_m}{\tau_c}\right))$$

The numerical data are accurately described by the exponential law throughout the investigated interval ($A \simeq 0.218, \tau_c \simeq 13.9$). The fitting parameter $\tau_c$ introduces naturally an **emergent geometric memory scale** that characterizes the transition from the rapid-growth regime to asymptotic saturation. Importantly, $\tau_c$ should not be interpreted as the physical memory time of the material. Instead, it is an emergent geometric quantity describing the characteristic scale over which the reconstructed transport geometry approaches its limiting state. It therefore constitutes a new emergent geometric observable generated by the reconstruction itself.

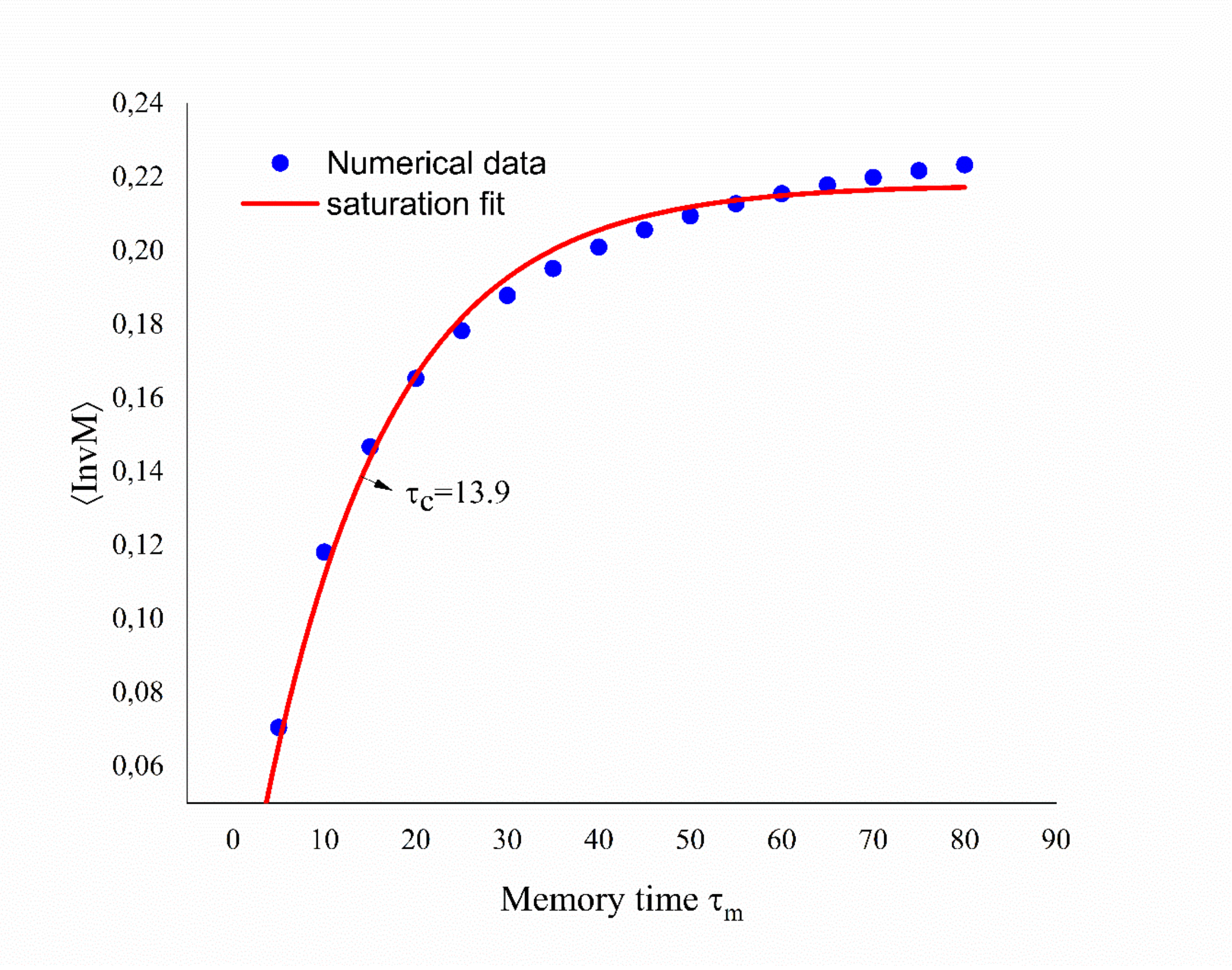

Fig 9-*Exponential fit of the averaged geometric invariant: The excellent agreement identifies the emergent geometric scale* ($\boldsymbol{\tau_c}$) *governing the crossover toward asymptotic saturation*

A coherent picture now emerges from Figs. 7–9 that complete the numerical validation of the universal geometric theory developed in the analytical sections. The simulations first confirm the predicted scaling collapse, then reveal the existence of an intrinsic saturation regime, and finally identify the emergent geometric scale ($\tau_c$) governing the crossover between rapid geometric accumulation and asymptotic evolution. These results demonstrate that finite-memory transport is organized by universal geometric laws extending far beyond the specific transport configurations considered here.

### 9.4 Geometric Susceptibility and Differential Signatures of Saturation

Having established the existence of an intrinsic saturation regime, we now examine the differential signatures that govern its emergence; the mechanism underlying this transition is more clearly revealed through differential measures of the geometric response.

To quantify the sensitivity of the reconstructed geometry to variations in memory time, we introduce the **geometric susceptibility**

$$\chi_m = \frac{d < I_m >}{d\tau_m}$$

whose numerical evolution is presented in Fig 10. The susceptibility is initially large, indicating that modest increases in memory produce substantial geometric amplification during the early stages of the reconstruction. As the memory time increases, however, the susceptibility decreases continuously, demonstrating that each additional increment of memory contributes progressively less to the geometric response than the previous one.

This behavior provides a direct differential signature of the transition toward the asymptotic regime identified in the previous subsection.

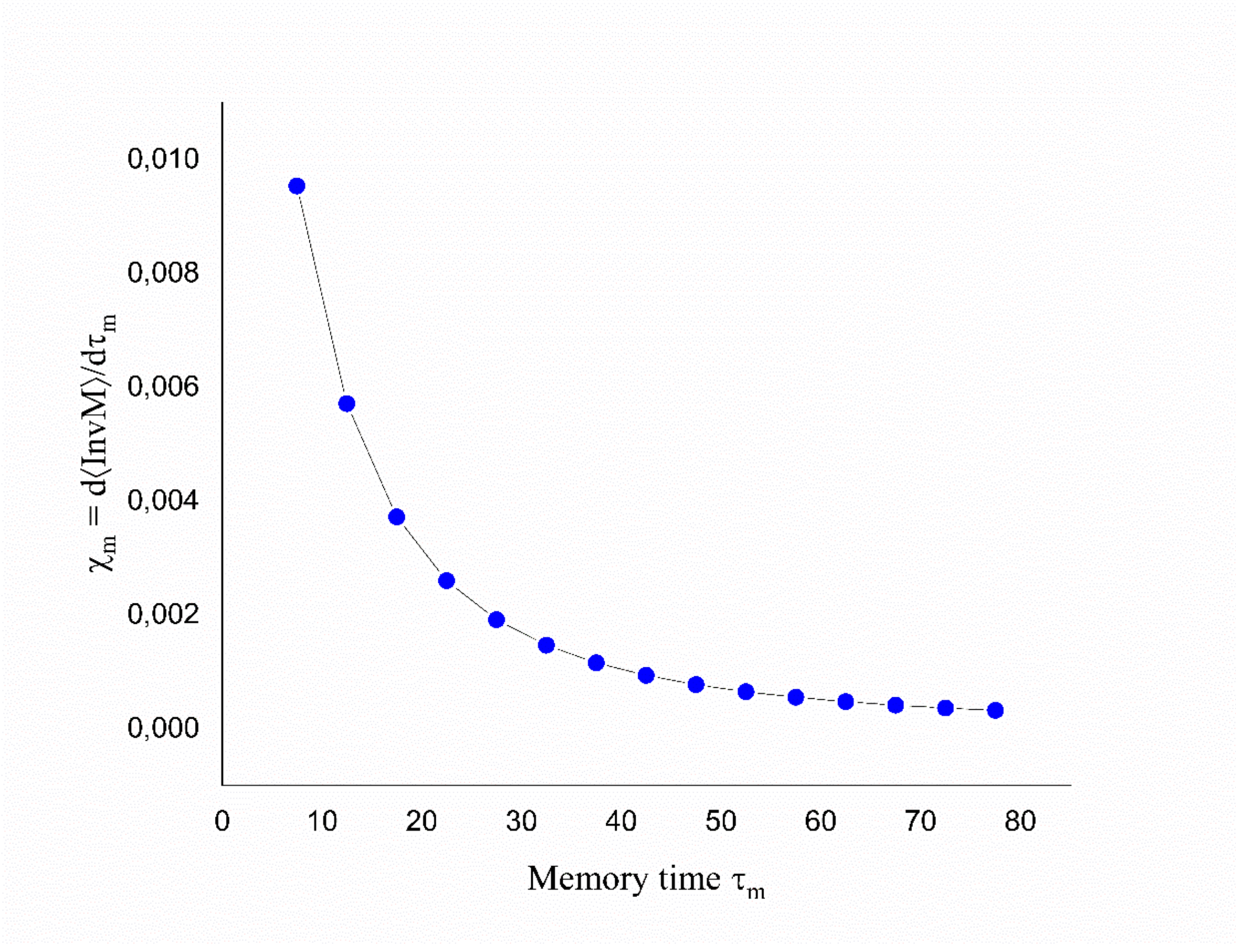

Fig 10-*Geometric susceptibility obtained from the first derivative of the averaged invariant with respect to memory time. The continuous decrease demonstrates the progressive reduction of the efficiency of geometric accumulation during the approach to saturation*

A complementary perspective is provided by the second derivative of the averaged geometric invariant, which characterizes the curvature of the growth law itself. Whereas the susceptibility measures the instantaneous efficiency of geometric accumulation, the second derivative quantifies how that efficiency evolves as the memory horizon expands.

The numerical results presented in Fig 11 show that the second derivative remains strictly negative over the entire investigated interval. The averaged geometric invariant is therefore globally concave, demonstrating that the efficiency of geometric accumulation decreases continuously during the reconstruction process. No inflection point, secondary acceleration, or reversal of concavity is observed.

This behavior has important physical consequences: The persistent global concavity rules out the existence of a delayed amplification stage, hidden instability, or internal phase transition governing the geometric evolution. Instead, the numerical results demonstrate that the reconstructed transport geometry evolves through a smooth and continuous crossover from rapid geometric accumulation to asymptotic saturation. The combined analysis of the geometric susceptibility and the second derivative therefore identifies a universal three-stage evolution of memory-induced transport:

- Growth regime ($\tau_m \ll \tau_c$), where geometric structure accumulates rapidly and the invariant increases rapidly.
- Crossover regime ($\tau_m \approx \tau_c$), where the efficiency of geometric accumulation decreases continuously as the reconstruction approaches its intrinsic geometric limit.
- Saturation regime ($\tau_m \gg \tau_c$), where additional memory produces only marginal additional geometric accumulation in the invariant.

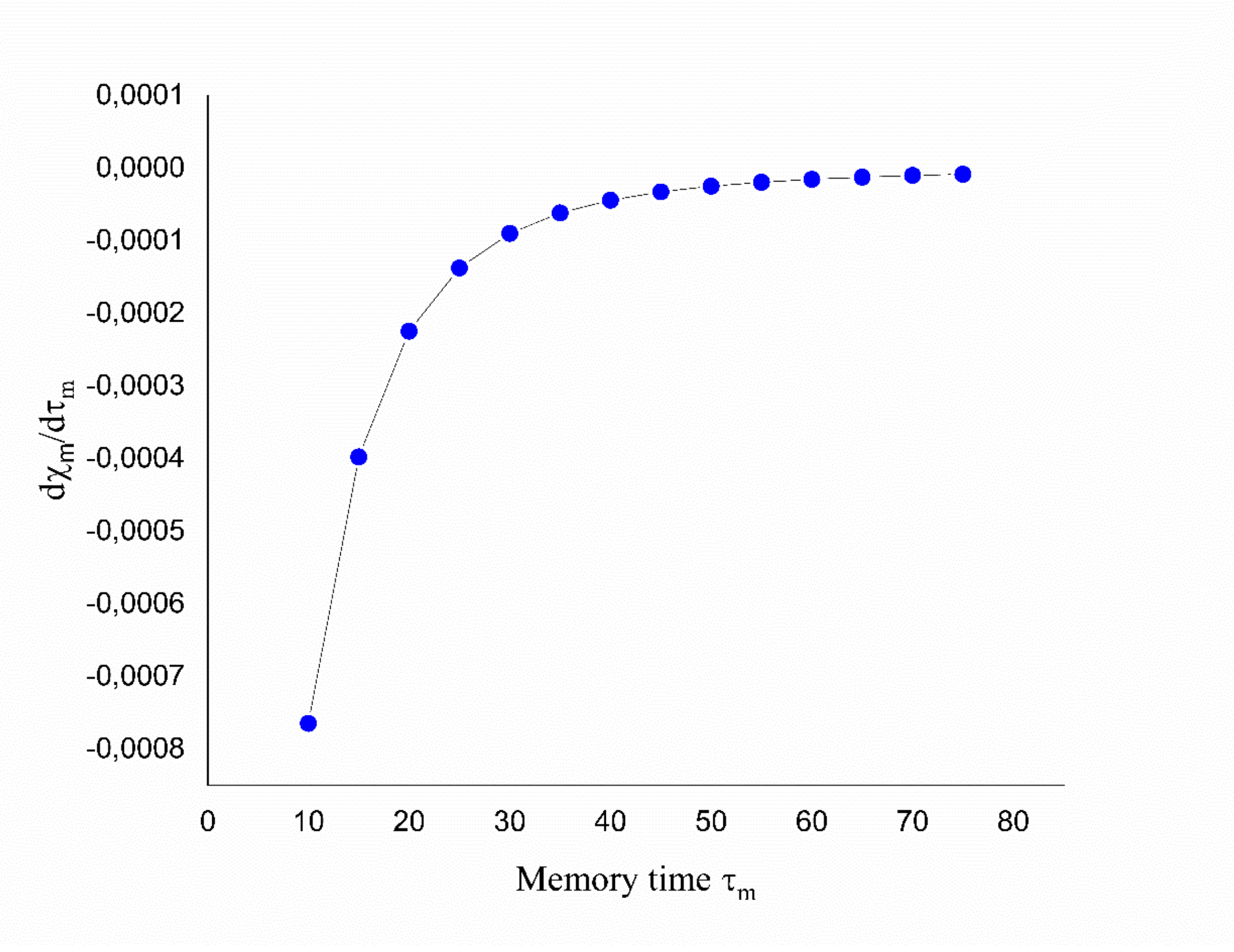

Fig 11- *Second derivative of the averaged geometric invariant: The smooth evolution confirms that geometric saturation results from a smooth geometric crossover in the reconstruction dynamics rather than from a discontinuous transition*

Altogether, Figs. 10 and 11 complete the differential characterization of memory-induced transport. The first derivative quantifies the instantaneous efficiency with which finite memory generates new geometric structure, whereas the second derivative reveals the continuous reduction of that efficiency as the reconstructed geometry approaches its asymptotic state. Combined with the local curvature fields, global invariants, and universal scaling laws established in the preceding subsections, these differential observables provide a complete numerical validation of the geometric framework developed throughout this work.

## 10 Discussion: Universal Geometry, Emergent Memory Scale, and the Prospect of Memory Resonance

The analytical developments and numerical results presented throughout this work converge toward a coherent geometric interpretation of finite-memory transport. Rather than acting as a constitutive correction to an otherwise local dynamics, finite memory reconstructs transport from causal histories, thereby generating a history-dependent connection whose ordered evolution naturally produces curvature, holonomy, and irreversible transport. The numerical investigation further reveals that this geometric mechanism possesses a well-defined internal organization characterized by universal scaling, an emergent characteristic memory scale, and a self-limited saturation regime.

### 10.1 Universal Geometry of Memory-Induced Transport

One of the central conclusions emerging from the present study is that the existence of memory-induced transport geometry appears to be largely insensitive to the detailed mathematical representation of the memory kernel. Throughout the theoretical construction, curvature arises as a direct consequence of two fundamental ingredients alone: finite causal memory and the temporal ordering of reconstructed transport operations. Once these conditions are satisfied, noncommutativity naturally develops, producing curvature and the associated holonomy responsible for irreversible transport.

More importantly, this observation suggests that the geometric sector of the theory possesses a remarkable degree of universality. The appearance of a memory-dependent connection, its associated curvature, and the resulting irreversible displacement should therefore not be regarded as consequences of a particular constitutive model or a specific kernel construction. Instead, they appear to represent generic geometric features of causal transport with finite memory. This conclusion is consistent with the general theoretical analysis developed previously, where the emergence of curvature was shown to depend primarily on causal reconstruction rather than on the detailed functional form of the relaxation kernel. Finite causal memory therefore appears sufficient to generate a nontrivial transport geometry.

### 10.2 Emergent Characteristic Memory Scale

Beyond demonstrating the existence of memory-induced geometry, the numerical analysis reveals the appearance of a second fundamental quantity: the characteristic memory scale $\tau_c \simeq 13.9$. Unlike the microscopic memory time $\tau_m$, which specifies the temporal horizon of causal reconstruction, $\tau_c$ emerges from the collective evolution of the reconstructed transport geometry itself. It therefore constitutes an **emergent geometric timescale** rather than an intrinsic material parameter.

Within the present dimensionless formulation, the corresponding physical value becomes $\tau_c^{phys} = \frac{\tau_c}{\omega}$, showing that its numerical value depends on the characteristic timescale of the physical system under consideration. Accordingly, the same geometric mechanism may operate over an enormous range of physical timescales, extending from ultrafast electronic and molecular processes to slowly evolving geological or astrophysical systems.

Its universality lies not in its numerical value but in its geometric role as the characteristic scale separating rapid geometric accumulation from the onset of saturation. The emergence of such an intrinsic geometric scale is particularly significant because it is not introduced phenomenologically. Instead, it arises naturally from the global organization of memory-induced transport revealed by the numerical simulations.

### 10.3 Geometric Origin of Saturation

The numerical analysis also clarifies the physical origin of the saturation regime observed for long memory times. The asymptotic approach of the geometric invariant does not indicate the disappearance of memory effects, nor does it imply that the reconstructed connection ceases to evolve. Instead, it reflects the progressive reduction in the efficiency with which additional transport history generates new geometric information. As the memory window becomes increasingly long, newly incorporated transport histories become increasingly correlated with those already contained within the reconstruction. Consequently, successive extensions of the memory horizon provide progressively less independent noncommutative information, causing the geometric invariant to approach a finite limiting value. The observed saturation should therefore be interpreted as an intrinsic property of causal reconstruction itself. Memory-induced geometry is intrinsically self-limiting: it continuously accumulates irreversible transport until the available transport history has been incorporated with maximal geometric efficiency.

### 10.4 Geometric Universality versus Dynamical Universality

The present results also suggest an important conceptual distinction between two different levels of universality. The first concerns the existence of the geometric mechanism itself: Memory-induced connection, curvature, and holonomy appear whenever transport is reconstructed from finite causal histories, independently of the detailed mathematical form of the memory kernel. This constitutes a universal geometric sector of the theory.

The second concerns the quantitative manifestation of geometric observables. While the existence of curvature appears universal, the detailed evolution of quantities such as the geometric invariant, susceptibility, or saturation characteristics may retain sensitivity to the internal structure of the memory kernel. Different kernels possessing multiple relaxation channels, broad relaxation spectra, or oscillatory components may therefore generate quantitatively distinct response functions while preserving the same underlying geometric mechanism. This distinction between a **universal geometric sector** and a **kernel-dependent dynamical sector** provides a natural framework for interpreting memory-induced transport across a broad range of physical systems. Geometry provides the universal mechanism responsible for irreversible transport, whereas the detailed response spectrum carries information about the microscopic architecture responsible for memory.

Moreover, this distinction echoes the broader perspective of nonequilibrium physics, where irreversibility emerges from the organization of temporal evolution rather than from equilibrium state variables alone, extending the thermodynamic perspective developed in nonequilibrium statistical physics toward a geometric description of transport irreversibility [12,14].

### 10.5 Toward Memory Resonance

An immediate implication of this distinction concerns the possibility of memory resonance. Within the framework investigated here, no resonance peak is observed. The geometric susceptibility decreases monotonically, while the second derivative remains negative throughout the investigated parameter range, demonstrating that the approach toward saturation is smooth and free of secondary amplification. This behavior is entirely consistent with the monotonic exponential relaxation kernel adopted throughout the present work.

However, the absence of resonance should not be regarded as a universal property of memory-induced geometry. Rather, it reflects the specific spectral structure of the kernel considered here. Memory kernels possessing intrinsic oscillatory components, competing relaxation processes, or multiple characteristic memory scales may generate a resonant amplification of geometric observables whenever their internal frequencies become commensurate with the external forcing frequency.

In such systems, the geometric invariant itself may exhibit resonant enhancement while remaining governed by the same geometric principles established in the present theory.

The possibility of memory resonance therefore constitutes a natural extension of the present framework. More generally, the spectral response of geometric observables may provide a powerful means of classifying memory kernels according to their microscopic dynamical organization while preserving the universal geometric structure generated by causal reconstruction.

In this perspective, finite memory is no longer viewed merely as a temporal correction acting on transport dynamics. It becomes the physical mechanism through which transport acquires an intrinsic geometric structure. Geometry provides the universal framework governing irreversible transport, whereas the spectral response of geometric observables retains the microscopic signature of the underlying memory processes. This distinction suggests a new conceptual route toward understanding irreversibility: **the geometry is universal, while the spectrum reveals the physics of memory**.

## 11 Conclusion

In this work, we have developed a geometric theory of finite-memory transport in which deformation is reconstructed from causal transport histories rather than prescribed through instantaneous local kinematics. Within this framework, finite memory acts as the mechanism through which transport acquires an intrinsic geometric structure-leading naturally to curvature, holonomy, and irreversible transport without invoking vorticity, stochastic forcing, constitutive nonlinearities, or explicit symmetry breaking.

The analytical and numerical results demonstrate that this memory-generated geometry possesses a coherent and universal organization. The intrinsic curvature invariant provides a coordinate-independent measure of the accumulated transport geometry, while the universal scaling laws and the emergence of a characteristic memory scale reveal that irreversible transport evolves through a well-defined geometric accumulation process before reaching an asymptotic saturation regime. Taken together, these results demonstrate that finite memory does not merely modify transport dynamics; it organizes the geometry of transport itself.

Beyond the specific mathematical construction developed here, the present framework suggests a broader conceptual distinction**.**

The emergence of transport geometry appears to be a universal consequence of causal memory, whereas the quantitative manifestation of geometric observables may retain information about the microscopic architecture of the underlying memory process. Geometry therefore provides the universal framework of memory-induced transport, whereas measurable geometric observables retain the physical signature of the underlying memory processes. Unlike conventional non-Markovian formulations, the present theory does not begin with prescribed transport equations to which memory is subsequently added. Instead, the transport connection itself is reconstructed from causal history, allowing geometry to emerge as a consequence of finite memory.

In this perspective, geometry is not imposed on transport but emerges from the causal organization of its history. We anticipate that the present framework can be extended to generalized memory kernels, multidimensional transport, turbulent and active flows, and a broad class of non-Markovian systems. More generally, the results presented here suggest that finite memory may constitute a universal geometric principle through which irreversible transport emerges across nonequilibrium physics. If confirmed in broader settings, this perspective would shift the role of memory from a constitutive ingredient of transport dynamics to a fundamental generator of transport geometry itself.

## 12 References

**[1]** L. D. Landau and E. M. Lifshitz, *Fluid Mechanics*, 2nd ed. (Pergamon Press, Oxford, 1987).

**[2]** G. K. Batchelor, *An Introduction to Fluid Dynamics* (Cambridge University Press, Cambridge, England, 1967).

**[3]** C. Truesdell and W. Noll, *The Non-Linear Field Theories of Mechanics*, 3rd ed. (Springer, Berlin, 2004).

**[4**] J. C. Maxwell, "On the Dynamical Theory of Gases," Philos. Trans. R. Soc. London 157, 49–88 (1867).

**[5]** R. S. Rivlin and J. L. Ericksen, "*Stress-Deformation Relations for Isotropic Materials*," J. Rational Mech. Anal. 4, 323–425 (1955).

**[6]** J. G. Oldroyd, "*On the Formulation of Rheological Equations of State*," Proc. R. Soc. London A 200, 523–541 (1950).

**[7]** R. B. Bird, R. C. Armstrong, and O. Hassager, *Dynamics of Polymeric Liquids*, Vol. 1, 2nd ed. (John Wiley & Sons, New York, 1987).

**[8]** H. Mori, "*Transport, Collective Motion, and Brownian Motion*," Prog. Theor. Phys. 33, 423 (1965).

**[9]** R. Zwanzig, "*Memory Effects in Irreversible Thermodynamics*," Phys. Rev. 124, 983 (1961).

**[10]** R. Zwanzig, *Nonequilibrium Statistical Mechanics* (Oxford University Press, New York, 2001).

**[11]** J. Keeling, E. M. Stoudenmire, M. C. Bañuls, and D. R. Reichman, *Process tensor approaches to non-Markovian quantum dynamics*, Phys. Rev. X 16, 020502 (2026).

**[12]** I. Prigogine, *From Being to Becoming: Time and Complexity in the Physical Sciences* (W. H. Freeman, San Francisco, 1980).

**[13]** I. Prigogine and I. Stengers, *Order Out of Chaos: Man's New Dialogue with Nature* (Bantam Books, New York, 1984).

**[14]** S. R. de Groot and P. Mazur, *Non-Equilibrium Thermodynamics* (Dover Publications, Mineola, NY, 1984).

**[15]** H.-P. Breuer, E.-M. Laine, J. Piilo, and B. Vacchini, "*Colloquium: Non-Markovian dynamics in open quantum systems*," Rev. Mod. Phys. 88, 021002 (2016).

**[16]** C. Lim and J. H. Jeon, "*Anomalous diffusion in coupled viscoelastic media: A fractional Langevin equation approach*," Phys. Rev. Research 7, 043356 (2025).

**[17]** T. Sandev, A. Iomin, J. Kurths, and L. Kocarev, *Shear-driven anomalous diffusion: Memory effects and stochastic resetting*, Phys. Fluids 37, 064101 (2025).

**[18]** S. Rijavec and G. Di Pietra, "*Tunable non-Markovian dynamics in a collision model: An application to coherent transport*," New J. Phys. 27, 043003 (2025).

**[19]** K. I. Mazzitello, D. G. Zarlenga, and C. M. Arizmendi, *Memory-induced transport and arrest in flashing ratchets*: From superdiffusion to clustering, Soft Matter **10**, 12 (2026).

**[20]** B. Vacchini, "*Generalized master equations leading to completely positive dynamics*," Phys. Rev. Lett. 117, 230401 (2016).

**[21]** T. Sandev, L. Kocarev, and R. Metzler, *Anomalous diffusion and fluctuations in complex systems and networks*, Chaos **36**, 013101 (2026).

**[22]** M. K. Wiśniewski, J. Łuczka, and J. M. Spiechowicz, *Effective mass approach to memory in non-Markovian systems*, Phys. Rev. E **109**, 044101 (2024).

**[23]** M. K. Wiśniewski, J. Łuczka, and J. M. Spiechowicz, *Memory corrections to Markovian Langevin dynamics*, Entropy **26**, 380 (2024).

**[24]** A. Sarkar, "*Non-Markovian route to coherence in heterogeneous diffusive systems*," *Phys. Rev. E* **112**, 054117 (2025).

**[25]** L. Lyu and H. Lei, "*Construction of coarse-grained molecular dynamics with many-body non-Markovian memory*," *Phys. Rev. Lett.* **131**, 177301 (2023).

**[26]** S. A. Loos and A. Godec, "*Non-Markovian effects in nonequilibrium systems*," *J. Phys. A: Math. Theor.* **58**, 220301 (2025).

**[27]** A. Abbasi, R. R. Netz, and A. Naji, "*Non-Markovian modeling of nonequilibrium fluctuations and dissipation in active viscoelastic biomatter*," *Phys. Rev. Lett.* **131**, 228202 (2023).

**[28]** L. Caprini *et al*., "*Emergent memory from tapping collisions in active granular matter*," *Commun. Phys.* **7**, 52 (2024).

**[29]** D. Andrieux, "*Geometric foundation of nonequilibrium transport: A Minkowski embedding of Markov dynamics*," *Phys. Rev. E* **111**, 054109 (2025).

**[30]** Z. Wang and J. Ren, "*Thermodynamic geometry of nonequilibrium fluctuations in cyclically driven transport*," *Phys. Rev. Lett.* **132**, 207101 (2024).

**[31]** C. Barbachoux and J. Kouneiher, "*Analytical and geometric foundations and modern applications of kinetic equations and optimal transport*," *Axioms* **14**, 350 (2025).

**[32]** A. Zhong and M. R. DeWeese, "*Beyond linear response: Equivalence between thermodynamic geometry and optimal transport*," *Phys. Rev. Lett.* **133**, 057102 (2024).

**[33]** M. Suárez-Rodríguez, F. De Juan, I. Souza, M. Gobbi, F. Casanova, and L. E. Hueso, "*Nonlinear transport in non-centrosymmetric systems*," *Nat. Mater.* **24**, 1005 (2025).

**[34]** S. Chennakesavalu and G. M. Rotskoff, "*Unified geometric framework for nonequilibrium protocol optimization*," *Phys. Rev. Lett.* **130**, 107101 (2023).

**[35]** D. Mandal, S. Sarkar, K. Das, and A. Agarwal, "*Quantum geometry induced third-order nonlinear transport responses*," *Phys. Rev. B* **110**, 195131 (2024).

**[36]** B. Doyon, S. Gopalakrishnan, F. Møller, J. Schmiedmayer, and R. Vasseur, "*Generalized hydrodynamics: A perspective*," *Phys. Rev. X* **15**, 010501 (2025).

**[37]** M. V. Berry, "*Quantal Phase Factors Accompanying Adiabatic Changes*," *Proc. R. Soc. London A* **392**, 45 (1984).

**[38]** J. H. Hannay, "*Angle Variable Holonomy in Adiabatic Excursion of an Integrable Hamiltonian*," *J. Phys. A: Math. Gen.* **18**, 221 (1985).

**[39]** A. Shapere and F. Wilczek, eds., *Geometric Phases in Physics* (World Scientific, Singapore, 1989).

**[40]** S. Kobayashi and K. Nomizu, *Foundations of Differential Geometry*, Vol. I (Wiley, New York, 1963).

**[41]** M. Nakahara, *Geometry, Topology and Physics*, 2nd ed. (Institute of Physics Publishing, Bristol, 2003).

**[42]** Kassmi, M., *Memory-induced curvature drives irreversible transport in irrotational flows*, arXiv:2604.08599 (2026).

(See below the Supplemental Material)

Supplemental Material

# Geometric Foundations of Memory-Generated Transport Geometry

## S1. Tensorial Formulation of Memory-Dependent Transport

A smooth differentiable manifold (M) that represents the configuration space of the carried material is used to formulate transport. Infinitesimal material displacements at each point $x \in M$ belong to the tangent space $T_xM$, and the tangent bundle is defined as the collection of all tangent spaces:

$$TM = \bigcup_{x \in M} T_xM$$

The causal reconstruction presented in the manuscript defines at each time a linear operator acting on the tangent bundle, $A_m(t): TM \to TM$ which will be referred to as the **memory-dependent transport endomorphism**. In a local coordinate map, it is represented by the matrix-valued field given by:

$$[A_m]^i{}_j(x,t) = \int_0^\infty K(\tau)\, \nabla_j u^i(x, t-\tau)\, d\tau$$

where $K(\tau)$ is a causal memory kernel satisfying $K(\tau) = 0,\ \tau < 0$, together with the normalization condition, $\int_0^\infty K(\tau)\, d\tau = 1$. No particular functional form is assumed beyond causality and the existence of a finite characteristic memory time. Unlike the instantaneous velocity-gradient operator of classical continuum mechanics, $A_m(t)$ is constructed from a finite transport history. Consequently, it should not be interpreted as an instantaneous deformation tensor but rather as a time-dependent bundle endomorphism generated by causal reconstruction.

The evolution of an infinitesimal material displacement $\delta x(t) \in T_{x(t)}M$ is governed by the linear evolution equation:

$$\frac{D\delta x(t)}{Dt} = A_m(t)\, \delta x(t)$$

Or, in local coordinates,

$$\delta \dot{x}^i(t) = (A_m)^i{}_j(t)\, \delta x^j(t)$$

In this case, the covariant notation highlights the fact that the transport operator operates on tangent vectors naturally, regardless of the coordinate system in which they are represented. The corresponding finite evolution defines a family of linear maps:

$$U(t, t_0): T_{x(t_0)}M \ \to T_{x(t)}M$$

Satisfying,

$$\delta x(t) = U(t, t_0)\, \delta x(t_0)$$

The propagator of transport itself satisfies a Cauchy-type differential equation,

$\dot{U}(t, t_0) = A_m(t)\, U(t, t_0)$ with initial condition, $U(t_0, t_0) = \ Id_{T_{x(t_0)}M}$ where $Id_{T_{x(t_0)}M}$ means the identity endomorphism of the tangent space. The unique solution is the time-ordered exponential:

$$U(t, t_0) = \mathcal{P} \exp\left(\int_{t_0}^{t} A_m(s)\, ds)\right)$$

where $\mathcal{P}$ denotes chronological ordering. The appearance of the ordering operator reflects the fact that the reconstructed endomorphisms generally fail to commute,

$$[A_m(t_1), A_m(t_2)] \neq \ 0$$

Thus, the finite transport operator cannot be reduced to the exponential of a single generator, and its evolution depends on the chronological composition of infinitesimal transport maps. This tensorial formulation provides the mathematical framework underlying the geometric construction developed in the main text.

Once transport is generated by a family of noncommuting bundle endomorphisms, curvature, holonomy, and the associated geometric invariants arise naturally from their ordered composition.

**S2. Tensorial Structure of Memory-Induced Curvature**

Let $A_m(t) \in \mathrm{End}(TM)$ denote the memory-dependent transport endomorphism introduced in S1. Since the transport propagator is generated by a time-dependent family of bundle endomorphisms, successive infinitesimal transport maps need not commute.

The primary geometric object encoding this noncommutativity is the curvature endomorphism $\mathcal{R}(t_1, t_2) \in \mathrm{End}(TM)$ defined via the commutator $[A_m(t_1), A_m(t_2)]$. In a local coordinate map, its components take the form:

$$\mathcal{R}^i{}_j(t_1, t_2) = [A_m]^i{}_k(t_1)[A_m]^k{}_j(t_2) - [A_m]^i{}_k(t_2)[A_m]^k{}_j(t_1)$$

Consequently, this curvature constitutes a smooth section of the endomorphism bundle $\mathrm{End}(TM)$, operating on tangent vectors in direct analogy to the reconstructed transport endomorphism. Unlike the classical Riemann curvature tensor, which is obtained from the commutator of covariant derivatives associated with an affine connection, the present object measures the noncommutativity of a time-dependent family of transport endomorphisms generated by causal reconstruction. Consequently, the two notions possess analogous algebraic structures while acting on different geometric objects. The transport propagator introduced in the previous section satisfies

$$U(t, t_0) = \mathcal{P} \exp\left(\int_{t_0}^{t} A_m(s)\, ds)\right)$$

whose logarithm admits the Magnus expansion

$$\log U = \Omega_1 + \Omega_2 + \cdots$$

The first correction beyond the averaged transport is

$$\Omega_2 = \frac{1}{2}\int_{t_0}^{t} dt_1 \int_{t_0}^{t_1} \mathcal{R}(t_1, t_2)\, dt_2$$

showing that the curvature endomorphism constitutes the first intrinsically noncommutative contribution to finite transport. This relation provides the mathematical origin of the geometric quantities introduced in the main manuscript: every higher-order correction generated by chronological ordering is ultimately constructed from successive commutators of the reconstructed transport endomorphism.

The tensorial character of the curvature follows immediately from that of $\mathrm{A_m}$. Under an arbitrary change of local frame, $\mathrm{A_m} \longmapsto \widetilde{\mathrm{A}}_\mathrm{m} = \mathrm{P^{-1}A_mP}$, where $\mathrm{P: M} \rightarrow \mathrm{GL(n)}$ is an invertible bundle automorphism, one obtains

$$\tilde{\mathcal{R}} = \left[\widetilde{\mathrm{A}}_\mathrm{m}(\mathrm{t_1}), \widetilde{\mathrm{A}}_\mathrm{m}(\mathrm{t_2})\right] = \mathrm{P}^{-1}\mathcal{R}\mathrm{P}$$

Thus, $\mathcal{R}$ transforms by conjugation and remains an intrinsic section of $\mathrm{End(TM)}$.

Its existence is therefore independent of the chosen local frame. In particular, $\mathcal{R}(\mathrm{t_1}, \mathrm{t_2}) = 0$, if and only if $[\mathrm{A_m}(\mathrm{t_1}), \mathrm{A_m}(\mathrm{t_2})] = 0$, in which case the reconstructed transport endomorphisms commute pairwise and the second Magnus contribution vanishes identically.

The finite transport operator is then completely determined by the first-order term of the Magnus expansion. The curvature endomorphism therefore provides the fundamental geometric object associated with causal transport reconstruction. Its algebraic properties form the basis for the construction of the intrinsic scalar invariants developed in the following section.

**S3. Spectral Geometry of the Memory-Induced Curvature**

The curvature endomorphism introduced in the previous section, $\mathcal{R}(\mathrm{t_1}, \mathrm{t_2}) \in \mathrm{End(TM)}$ is intrinsically defined on the tangent bundle and therefore exists independently of any particular coordinate representation. Its local matrix components depend on the chosen frame, whereas its algebraic spectrum is an intrinsic property of the endomorphism itself. This naturally leads to the construction of scalar quantities that remain invariant under arbitrary bundle automorphisms. The action of the general linear group on the endomorphism bundle is given by conjugation,

$$\mathcal{R} \longmapsto \tilde{\mathcal{R}} = \mathrm{P}^{-1}\mathcal{R}\mathrm{P}, \qquad \mathrm{P: M} \rightarrow \mathrm{GL(n)}$$

so that intrinsic scalar quantities are naturally identified with conjugacy invariants of the curvature endomorphism. Among the infinite family of such invariants, the present work introduces the quadratic spectral invariant $\mathrm{I_m} = \mathrm{Tr}(\mathcal{R}^2)$, which constitutes the fundamental intrinsic scalar employed throughout the analytical and numerical developments of the main manuscript. Under the induced action of the bundle automorphism, $\tilde{\mathcal{R}}^2 = \mathrm{P}^{-1}\mathcal{R}^2\mathrm{P}$, and the cyclic property of the trace immediately yields

$$\mathrm{Tr}\left(\tilde{\mathcal{R}}^2\right) = \mathrm{Tr}(\mathrm{P}^{-1}\mathcal{R}^2\mathrm{P}) = \mathrm{Tr}\,(\mathcal{R}^2)$$

Hence, $\mathrm{I_m} = \mathrm{Tr}(\mathcal{R}^2)$ is invariant under arbitrary changes of local frame and therefore defines an intrinsic scalar of the reconstructed transport geometry. This invariance is substantially stronger than orthogonal invariance. For instance, $\mathrm{Tr}(\mathcal{R}^{\mathrm{T}}\mathcal{R})$ remains invariant only under orthogonal transformations, $\mathrm{P} \in \mathrm{O(n)}$, whereas $\mathrm{Tr}(\mathcal{R}^2)$ is preserved under the full action of $\mathrm{GL(n)}$. Its value therefore depends exclusively on the similarity class of the curvature endomorphism and not on the particular representation adopted for the transport dynamics.

More generally, every polynomial function $\mathrm{Tr}\left(\mathcal{R}^{\mathrm{k}}\right), \mathrm{k} = 1,2, \ldots$, defines a spectral invariant of the memory-induced curvature. These invariants belong to the classical invariant theory of linear endomorphisms and characterize the intrinsic algebraic structure of the reconstructed transport connection.

The contribution of the present work is not the introduction of these invariants themselves, which are classical objects in differential geometry and representation theory. Rather, it is the identification of the memory-induced curvature endomorphism as the geometric object from which such invariants can be constructed, together with the demonstration that they acquire a direct physical interpretation in finite-memory transport. In particular, the invariant $\mathrm{I_m}$ becomes the intrinsic geometric quantity governing the universal scaling laws and numerical observables derived in the main manuscript. This establishes a direct bridge between classical spectral geometry and the irreversible transport generated by causal memory reconstruction.

Unlike the matrix components of the curvature, which vary with the local representation, the spectral invariant $\mathrm{I_m}$ characterizes an intrinsic property of the reconstructed transport geometry itself and therefore provides the natural scalar descriptor of the geometric framework developed in the present work.

## S4. Energy Functionals on the Memory-Induced Transport Geometry

The previous sections established that finite-memory transport naturally generates a bundle endomorphism $A_m(t) \in \Gamma(\mathrm{End}(TM))$, whose ordered evolution gives rise to the curvature endomorphism $\mathcal{R}(t_1, t_2) \in \Gamma(\mathrm{End}(TM))$, and to the associated spectral invariant:

$$I_m = \mathrm{Tr}(\mathcal{R}^2)$$

These objects characterize the intrinsic geometry generated by causal transport reconstruction independently of any transported material quantity. A complementary question is whether this geometric structure also induces natural energetic functionals. More precisely, given the ordered evolution generated by the memory connection, one may ask whether the accumulated transport admits canonical scalar quadratic forms analogous to those arising in Riemannian geometry, elasticity, or gauge theory. The objective of the present section is therefore not to introduce a phenomenological transport energy, but rather to identify the natural quadratic functionals canonically associated with the geometry generated by finite memory.

### 4.1 Canonical Quadratic Energy Functional

The leading geometric contribution to the ordered evolution is represented by the second Magnus operator $\Omega_2 = \frac{1}{2}\int_{t_0}^{t} dt_1 \int_{t_0}^{t_1} \mathcal{R}(t_1, t_2)\, dt_2$, which belongs to $\Gamma(\mathrm{End}(TM))$.

Since $\Omega_2$ is itself a bundle endomorphism, every tangent vector $\Delta x \in T_pM$, naturally determines a quadratic scalar through the metric tensor (g):

$$E_R(\Delta x) = g(\Delta x, \Omega_2 \Delta x)$$

In local coordinates,

$$E_R = g_{ij}\, \Delta x^i\, (\Omega_2)^j{}_k \Delta x^k$$

This expression defines the canonical quadratic functional associated with the memory-generated transport geometry. Unlike conventional transport energies constructed from velocities or constitutive stresses, the present functional depends exclusively on the ordered evolution generated by the reconstructed connection.

Its existence follows directly from the differential-geometric structure established in the previous sections and requires no additional constitutive assumptions.

### 4.2 Coordinate Independence

Because the metric tensor performs a complete contraction of all tensorial indices, the functional $E_R$ is an intrinsic scalar on the tangent bundle.

Indeed, under an arbitrary coordinate transformation, $x^i \to x^{i'}$, the metric, tangent vector, and bundle endomorphism transform according to the standard tensorial rules,

$$g \longmapsto g', \Delta x \longmapsto \Delta x',\ \Omega_2 \longmapsto\ \Omega_2{}'$$

from which one immediately obtains

$$E'_R = g'(\Delta x', \Omega_2'\Delta x') =\ g(\Delta x, \Omega_2 \Delta x) = E_R$$

The geometric energy functional is therefore independent of the coordinate system and depends only on the intrinsic transport geometry generated by finite memory. For the Euclidean metric adopted in the numerical simulations, $g_{ij} = \delta_{ij}$, the general expression reduces to the familiar matrix form, $E_R = \Delta x^T \Omega_2 \Delta x$, used throughout the analytical developments of the main manuscript.

### 4.3 Local Energy Density

The quadratic functional admits a natural field-theoretic extension. Let $\xi \in \Gamma(TM)$ be a smooth displacement field defined over the transport manifold. The local geometric energy density is then $\varepsilon_R = g(\xi, \Omega_2 \xi)$ or, equivalently, $\varepsilon_R =\ g_{ij}\ \xi^i\ (\ \Omega_2)^j{}_k \xi^k$. The corresponding global functional is obtained by integration over the Riemannian volume element,

$E_R =\ \int_M \varepsilon_R\, dV_g$, where $dV_g =\ \sqrt{\ |g\ |}\, dx^1 \ldots dx^n$ denotes the canonical volume form associated with the metric. This formulation identifies the energetic contribution of memory-induced geometry as a global functional on the transport manifold rather than as a property of individual trajectories.

### 4.4 Relation with the Spectral Curvature Invariant

Although both $I_m = \mathrm{Tr}(\mathcal{R}^2)$ and $E_R = g(\Delta x, \Omega_2 \Delta x)$ originate from the same curvature endomorphism, they characterize fundamentally different geometric objects. The spectral invariant

$$I_m : \Gamma(\mathrm{End}(TM)) \rightarrow \mathbb{R}$$

depends exclusively on the spectrum of the curvature operator. It therefore measures an intrinsic property of the reconstructed geometry independently of any transported configuration. By contrast,

$$E_R : TM \times \Gamma(\mathrm{End}(TM)) \rightarrow \mathbb{R}$$

couples a tangent vector to the accumulated curvature through the metric tensor. Its value consequently depends on both the intrinsic geometry and the transported state.

The two constructions therefore play complementary roles: The invariant $I_m$ quantifies the intrinsic intensity of the memory-generated geometry itself, whereas the quadratic functional $E_R$ measures how that geometry acts on transported material displacements. This distinction mirrors the general framework of differential geometry, where curvature characterizes the manifold independently of the motion of particles, while energetic functionals describe the interaction between geometry and transported objects.

### 4.5 Differential Memory Response

Since the accumulated operator $\Omega_2$ depends parametrically on the characteristic memory time $\tau_m$, the quadratic functional inherits the same dependence. Its first variation with respect to memory is therefore

$$\frac{dE_{\mathcal{R}}}{d\tau_m} = g(\Delta x, \frac{d\,\Omega_2}{d\tau_m}\,\Delta x)$$

which defines the first geometric response of the energetic functional to variations of the reconstruction horizon. Higher-order variations,

$$\frac{d^n E_{\mathcal{R}}}{d^n \tau_m}$$

generate a hierarchy of memory-response functionals that characterize the progressive evolution of the energetic consequences of causal reconstruction.

Although these higher-order quantities are not investigated quantitatively in the present work, they provide a natural mathematical framework for future studies of memory-controlled transport and geometric transitions associated with complex memory kernels.

The construction developed in this section completes the geometric hierarchy associated with finite-memory transport. The memory kernel generates a connection on the tangent bundle, whose ordered evolution defines a curvature endomorphism. This curvature possesses intrinsic spectral invariants while simultaneously inducing canonical quadratic energy functionals through the metric structure of the transport manifold. Together, these objects constitute the fundamental geometric ingredients employed throughout the analytical developments of the present theory and establish that finite memory generates not only an intrinsic transport geometry, but also the natural energetic functionals associated with that geometry.

$$K(\tau) \rightarrow A_m \rightarrow \mathcal{R} \rightarrow \{I_m , E_R \} \rightarrow \Delta\gamma$$

**S5. Minimal Structural Conditions for Memory-Induced Curvature**

The preceding sections established that finite memory defines a bundle endomorphism $A_m(t) \in \Gamma(\mathrm{End}(TM))$, whose ordered evolution generates the curvature operator $\mathcal{R}(t_1, t_2) = [A_m(t_1), A_m(t_2)]$. A natural mathematical question is whether the existence of finite memory alone is sufficient to produce a nontrivial curvature. The following theorem shows that this is not the case.

a. **Theorem (Single-Mode Transport)**

Assume that the velocity-gradient field possesses the separable form $\nabla u(t) = A\, f(t)$, where $A \in \Gamma(\mathrm{End}(TM))$, is time independent and $f: \mathbb{R} \rightarrow \mathbb{R}$ is an arbitrary scalar function. Then, for every causal memory kernel $\mathcal{K}$, $\mathcal{R}(t_1, t_2) \equiv 0$

b. **Proof**

The reconstructed transport connection is $A_m(t) = \int_0^\infty K(s)\, A\, f(t - s)\, ds$. Since A is independent of time, $A_m(t) = A\, g(t)$, with $g(t) = \int_0^\infty K(s)\, f(t - s)\, ds$. Hence

$$A_m(t) \in \text{span}\,\{A\} \subset \Gamma(\text{End}(TM))$$

The image of the reconstruction therefore lies inside the one-dimensional commutative subalgebra generated by A. Consequently,

$$\mathcal{R}(t_1, t_2) = [A_m(t_1), A_m(t_2)] = g(t_1)g(t_2)[A, A] = g(t_1)g(t_2)(A^2 - A^2) = 0, \qquad \forall\ (t_i, t_j)$$

As a result, the second Magnus operator $\Omega_2$ vanishes, leading directly to a zero holonomy contribution.

### c. Corollary (Necessity of Noncommuting Deformation Modes)

Suppose that,

$$\nabla u(t) = \sum_{\alpha=1}^{N} A_\alpha\, f_\alpha(t)$$

where the endomorphisms $A_\alpha \in \Gamma(\text{End}(TM))$ are time independent. Then,

$$A_m(t) = \sum_{\alpha=1}^{N} A_\alpha\, g_\alpha(t), \quad g_\alpha(t) = \int_0^\infty K(s)\, f_\alpha(t-s) ds$$

The curvature becomes:

$$\mathcal{R}(t_1, t_2) = \sum_{\alpha,\beta} g_\alpha(t_1)\, g_\beta(t_1)\, [A_\alpha, A_\beta]$$

Therefore, $\mathcal{R} \equiv 0 \leftrightarrow \left[A_\alpha, A_\beta\right] = 0\,, \forall\ (\alpha, \beta)$

### d. Remark

The previous theorem separates two mathematically independent ingredients. The memory kernel determines the temporal reconstruction through the scalar functions $g_\alpha(t)$, whereas the existence of curvature depends exclusively on the Lie algebra generated by the family of bundle endomorphisms $\{A_\alpha\}$. Finite memory alone therefore cannot produce non-integrable transport geometry. Nontrivial curvature appears only when the reconstructed evolution explores non-Abelian subalgebra of $\Gamma(\text{End}(TM))$.

#### e. Consequence

The geometric mechanism developed throughout the present work may therefore be expressed as the chain:

**Finite Memory → Temporal Reconstruction → Non-Abelian Endomorphism Algebra → Curvature → Holonomy → Irreversible Transport**

Unlike the magnitude of the curvature, which depends on the temporal reconstruction induced by the memory kernel, its very existence is governed entirely by the algebraic structure of the reconstructed transport operators. This theorem therefore identifies the minimal structural conditions under which finite memory acquires geometric significance.

### S6. Computational Realization of the Geometric Framework

The analytical developments presented in the main manuscript define the memory-generated transport geometry through the sequence: $A_m \to \mathcal{R} \to I_m \to \Delta\gamma$

The purpose of the numerical procedure is therefore not merely to evaluate these quantities, but to construct discrete approximations of the underlying geometric objects and to verify the analytical predictions derived from the continuous theory. Throughout the computations, the numerical discretization is regarded as an approximation of the continuous transport-history manifold introduced in the theoretical formulation.

#### 6.1 Discretization of the Transport-History Manifold

The transport-history manifold $\mathcal{H} = [0, T] \times [0, T]$ is discretized by a uniform Cartesian grid,

$$\mathcal{H}_h = \{(t_i, t_j)\}_{i,j}^{N}$$

where,

$$t_i = i\Delta t, \qquad \Delta t = \frac{T}{N-1}$$

Each node of the computational grid represents one ordered pair of transport times $(t_1, t_1)$, allowing all geometric quantities to be evaluated over the complete history manifold.

The discretization therefore provides a finite-dimensional representation of the continuous geometry introduced in Sections 1–5.

### 6.2 Numerical Reconstruction of the Memory Connection

At every sampling time, the reconstructed transport connection is obtained by discretizing the causal convolution $A_m(t)$. The integral is approximated by a quadrature over the discrete memory history:

$$A_m(t_i) \approx \sum_k \mathcal{K}(s_k)\,\nabla(t_i - s_k)\Delta s$$

This reconstruction defines the discrete family of bundle endomorphisms

$A_m(t_i) \in \mathrm{End}(T_xM)$, which constitutes the fundamental numerical object from which every subsequent geometric observable is computed.

### 6.3 Numerical Curvature Operator

Once the reconstructed connection has been evaluated over the history manifold, the discrete curvature operator is obtained directly from the commutator,

$$\mathcal{R}(t_i, t_j) = [A_m(t_i), A_m(t_j)]$$

The numerical simulations therefore compute the curvature exactly from its defining algebraic relation, without introducing additional constitutive approximations. For graphical visualization, the matrix-valued curvature is represented through an effective scalar magnitude, $\mathcal{R}_{eff}(t_i, t_j)$ which preserves the spatial organization of the reconstructed geometry while allowing direct visualization over the transport-history manifold.

### 6.4 Spectral Geometric Observables

The principal numerical observable is the spectral curvature invariant $I_m = \mathrm{Tr}(\mathcal{R}^2)$, evaluated independently at every point of the discrete history manifold, $I_m(t_i, t_j)$. Unlike the curvature itself, this quantity is invariant under similarity transformations and therefore provides a numerical characterization of the intrinsic geometry independently of the chosen representation. Several global observables are subsequently constructed. The history-average is defined by:

$$< I_m > = \frac{1}{T^2} \iint_{\mathcal{H}} I_m(t_1, t_2)\, dt_1 dt_2$$

whose numerical approximation is obtained by replacing the integral with the corresponding discrete double sum.

To quantify the cumulative geometric activity generated as the memory horizon increases, we further introduce $I(\tau_m) = \int_0^{\tau_m} < I_m > d\tau$, which measures the accumulated intrinsic geometry produced during the reconstruction process. Finally, the memory-generated contribution to curvature is isolated through, $\Delta\mathcal{R} = \mathcal{R}(\tau_m) - \mathcal{R}(\tau_{ref})$, where the reference state corresponds to the smallest memory considered in the numerical analysis.

### 6.5 Differential Response Functions

The dependence of the reconstructed geometry on the characteristic memory time is characterized through successive derivatives with respect to $\tau_m$. The local geometric sensitivity is defined as $S_R = (\partial R/\partial \tau_m)$, and is evaluated numerically using centered finite differences. At the global level, the geometric susceptibility is defined by $\chi = \frac{d<I_m>}{d\tau_m}$, while the second derivative, $\frac{d^2<I_m>}{d^2\tau_m}$ characterizes the evolution of the amplification efficiency during memory reconstruction. These differential observables provide jointly a quantitative signatures of the transition from efficient geometric accumulation to the asymptotic saturation regime identified in the main manuscript.

### 6.6 Dimensionless Scaling Analysis

To compare transport geometries generated under different physical conditions, all numerical observables are represented as functions of the dimensionless control parameter $\omega\tau_m$.

This normalization removes the dependence on the individual forcing frequency and memory time, allowing the universal scaling predicted analytically to be examined directly. Within this representation, the numerical simulations reveal the three transport regimes discussed in the main text: **weak-memory regime**, **intermediate geometric crossover** and **long-memory saturation**.

### 6.7 Computational Validation Strategy

The numerical procedure mirrors the mathematical hierarchy established throughout the theoretical development. Beginning with the reconstruction of the memory connection, successive computations generate the curvature operator, its spectral invariant, the associated global observables, and finally the differential response functions governing their evolution with memory. Rather than constituting independent numerical diagnostics, these quantities represent successive geometric projections of the same underlying structure. The numerical analysis therefore provides a direct computational realization of the differential-geometric framework developed in the present work, allowing each analytical prediction to be examined through intrinsically defined geometric observables.

## S7. Mathematical Consistency and Geometric Generality

The geometric framework developed throughout this Supplemental Material is constructed as an extension of classical continuum transport rather than an alternative kinematic theory. Consequently, its mathematical formulation must satisfy two fundamental consistency requirements. First, the theory must recover the classical local description in the instantaneous-memory limit. Second, every geometric object introduced in the construction should possess an intrinsic meaning independent of the particular Euclidean representation employed in the numerical implementation. The present section establishes these two properties.

### 7.1 Classical Limit

The reconstructed transport connection is defined as the causal bundle endomorphism

$$A_m(t) = \int_0^\infty K(s)\,\nabla u(t-s)\,ds$$

Consider a family of admissible memory kernels $\mathcal{K}_{\tau_m}(s)$ parameterized by the characteristic memory time ($\tau_m$). Assume that $\mathcal{K}_{\tau_m} \to \delta,\ (\tau_m \to 0)$ in the sense of distributions.

The reconstruction operator then converges strongly to the instantaneous velocity-gradient field, $A_m(t) \to \nabla u(t)$. Since the transport evolution is now generated by a single local operator, chronological ordering ceases to introduce additional geometric information.

The curvature operator therefore satisfies $\mathcal{R}(t_1, t_2) \to 0$. Consequently, $\Omega_2 \to 0$, and every higher geometric quantity generated by the curvature vanishes continuously, $I_m \longrightarrow 0$ , $S_m \longrightarrow 0, \Delta\gamma \to 0$

Hence the complete geometric hierarchy contracts smoothly to the classical continuum description,

$$\tau_m \to 0 \Longrightarrow A_m \longrightarrow \nabla u \Longrightarrow \mathcal{R} \longrightarrow 0 \Longrightarrow \Omega_2 \to 0 \Longrightarrow I_m \longrightarrow 0 \Longrightarrow \Delta\gamma \to 0$$

No singular behavior appears during this limit, demonstrating that the present construction extends rather than replaces classical transport kinematics.

### 7.2 Independence from the Euclidean Representation

Although every numerical example presented in the main manuscript is performed in Euclidean transport space $(M, g) = (\mathbb{R}^n, \delta)$, none of the mathematical constructions developed in this Supplemental Material depends on this particular choice.

The reconstructed connection is defined intrinsically as a section of $\Gamma(\mathrm{End}(TM))$, the curvature is obtained from the commutator in the endomorphism algebra,

$\mathcal{R}(t_1, t_2) = [A_m(t_1), A_m(t_2)]$, and the principal geometric invariant is the spectral quantity $I_m = \mathrm{Tr}(\mathcal{R}^2)$, whose definition requires neither Cartesian coordinates nor a Euclidean metric. Likewise, the tensorial transport energy remains naturally defined by $E_R = g(\Delta x, \Omega_2 \Delta x)$, which is valid on every Riemannian manifold endowed with metric (g).

The Euclidean expressions employed throughout the numerical implementation therefore represent only one particular realization of a coordinate-free geometric construction.

### 7.3 Geometric Scope of the Formalism

The mathematical framework developed in the present work is therefore not restricted to transport in flat Euclidean space. Because every fundamental object is formulated in terms of bundle endomorphisms, operator commutators, tensor contractions, and spectral invariants, the construction extends naturally to arbitrary differentiable manifolds carrying a tangent bundle. Within such a setting, two independent geometric structures may coexist. The first is the background geometry associated with the manifold itself through its metric and affine connection. The second is the memory-generated geometry introduced in the present work through the causal reconstruction of transport histories.

These two geometric sectors are conceptually independent. The former describes the geometry of the underlying configuration space, whereas the latter characterizes the geometry generated by the transport dynamics. Nothing in the present formulation requires these two structures to coincide.

### 7.4 Perspective

The present theory therefore suggests a broader mathematical viewpoint. Rather than prescribing the transport connection from the outset, the connection is reconstructed from causal histories, after which curvature, spectral invariants, holonomy, and transport energetics emerge as intrinsic consequences of this reconstruction. This raises the possibility that increasingly general transport theories could be formulated in which part of the effective differential geometry is generated dynamically by memory itself. Whether such a construction may ultimately extend beyond the connection to influence the effective transport metric or more general geometric structures remains an open mathematical problem.

Its investigation lies beyond the scope of the present work, but the bundle-theoretic formulation developed throughout this Supplemental Material provides a natural starting point for such extensions.